%% file: libbi_arxiv.tex
\documentclass[english,article,a4paper,10pt]{article}
\usepackage[T1]{fontenc}
\usepackage[latin9]{inputenc}
\usepackage{listings}
\usepackage{verbatim}
\usepackage{float}
\usepackage{url}
\usepackage{amsmath}
\usepackage{amssymb}
\usepackage{graphicx}
\usepackage[numbers]{natbib}

\makeatletter

\providecommand{\tabularnewline}{\\}
\floatstyle{ruled}
\newfloat{algorithm}{tbp}{loa}
\providecommand{\algorithmname}{Algorithm}
\floatname{algorithm}{\protect\algorithmname}

\usepackage{fullpage}
\usepackage{times}
\newcommand{\pkg}{}
\newcommand{\proglang}{}
\author{Lawrence M. Murray\\
CSIRO Mathematics, Informatics and Statistics}
\date{}

\title{Bayesian State-Space Modelling on High-Performance Hardware Using \pkg{LibBi}}

\usepackage{float}
\usepackage{lmodern}
\usepackage{clrscode3e}
\usepackage{color}
\usepackage{alltt}
\usepackage[T1]{fontenc}

\input{libbi-highlight.sty}
\input{libbi.sty}

\graphicspath{{figs/}}

\floatstyle{ruled}
\newfloat{code}{tpbh}{aux}
\floatname{code}{Code}

\floatstyle{ruled}
\newfloat{file}{tpbh}{aux}
\floatname{file}{File}

\newcommand{\Foreach}{\kw{foreach}\,\Indentmore}

\renewcommand{\If}{\kw{if}\,\Indentmore}

\renewcommand{\For}{\kw{for}\,\Indentmore}

\makeatother

\usepackage{babel}
\begin{document}
\maketitle
\begin{abstract}
\pkg{LibBi} is a software package for state-space modelling and Bayesian
inference on modern computer hardware, including multi-core central
processing units (CPUs), many-core graphics processing units (GPUs)
and distributed-memory clusters of such devices. The software parses
a domain-specific language for model specification, then optimises,
generates, compiles and runs code for the given model, inference method
and hardware platform. In presenting the software, this work serves
as an introduction to state-space models and the specialised methods
developed for Bayesian inference with them. The focus is on sequential
Monte Carlo (SMC) methods such as the particle filter for state estimation,
and the particle Markov chain Monte Carlo (PMCMC) and SMC$^{2}$ methods
for parameter estimation. All are well-suited to current computer
hardware. Two examples are given and developed throughout, one a linear
three-element windkessel model of the human arterial system, the other
a nonlinear Lorenz '96 model. These are specified in the prescribed
modelling language, and \pkg{LibBi} demonstrated by performing inference
with them. Empirical results are presented, including a performance
comparison of the software with different hardware configurations.
\end{abstract}

\section{Introduction}

State-space models (SSMs) have important applications in the study
of physical, chemical and biological processes. Examples are numerous,
but include marine biogeochemistry~\citep{Dowd2006,Jones2010,Dowd2011,Parslow2012},
ecological population dynamics~\citep{Wikle2003,Newman2009,Peters2010,Hosack2012},
Functional Magnetic Resonance Imaging~\citep{Riera2004,Murray2008,Murray2011b},
biochemistry \citep{Golightly2008,Golightly2011} and object tracking~\citep{Vo2006}.
They are particularly useful for modelling uncertainties in the parameters,
states and observations of such processes, and particularly successful
where a rigorous probabilistic treatment of these uncertainties leads
to improved scientific insight or risk-informed decision making.

SSMs can be interpreted as a special class within the more general
class of Bayesian hierarchical models (BHMs). For a given data set,
existing methods for inference with BHMs, such as Gibbs sampling,
can be applied. However, specialist machinery for SSMs has been developed
to better deal with peculiarities of models in the class. These include
nonlinearity, multiple modality, missing closed-form densities and
significant correlations between state variables and parameters. Variants
of sequential Monte Carlo (SMC)~\citep{Doucet2001} are particularly
attractive, including particle Markov chain Monte Carlo (PMCMC)~\citep{Andrieu2010}
and SMC$^{2}$~\citep{Chopin2012}. These methods have two particularly
pragmatic qualities: they admit a wide range of SSMs, including nonlinear
and non-Gaussian models, without approximation bias, and they are
well-suited to recent, highly parallel, computer architectures~\citep{Lee2010,Murray2013}.

Commodity computing hardware has diverged from the monoculture of
x86 CPUs in the 1990s to the ecosystem of diverse desktop central
processing units (CPUs), mobile processors and specialist graphics
processing units (GPUs) of today. Adding to the challenge since 2004
is that Moore's Law, as applied to computing performance, has been
upheld not by increasing clock speed, but by broadening parallelism~\citep{Sutter2005}.
This should not be understood as a passing fad, nor a deliberate design
choice for modern applications, but as a necessity enforced by physical
limits, the most critical of which is energy consumption~\citep{Ross2008}.
Thus, while architectures continue to change, their reliance on parallelism
is unlikely to, at least in the foreseeable future. The implication
for statistical computing is clear: in order to make best use of current
and future architectures, algorithms must be parallelised. On this
criterion, SMC methods are a good fit.

Given these specialised methods for SSMs, it is appropriate that specialised
software be available also. \pkg{LibBi}%
\footnote{\url{http://www.libbi.org}%
} is such a package. Nominally, the name is a contraction of ``Library
for Bayesian inference'', and pronounced ``Libby''. Its design
goals are accessibility and speed. It accepts SSMs specified in its
own domain-specific modelling language, which is parsed and optimised
to generate and compile \proglang{C++} code for the execution of
inference methods. The code exploits technologies that include SSE
for vector parallelism, OpenMP for multithreaded shared-memory parallelism,
MPI for distributed-memory parallelism, and CUDA for GPU parallelism.
The user interacts with the package via a command-line interface,
with input and output files handled in the standard NetCDF format,
based on the high-performance HDF5.

This work serves as a brief introduction to SSMs, appropriate Bayesian
inference methods for them, the modern computing context, and the
consideration of all three in \pkg{LibBi}. The material is presented
in three parts: Section \ref{sec:models} introduces SSMs as a special
class of BHMs; Section \ref{sec:methods} provides specialist methods
for inference with the class; Section \ref{sec:software} provides
technical information on \pkg{LibBi} itself. Two examples are developed
throughout. At the end of Section \ref{sec:models}, these examples
are developed as SSMs and specified in the \pkg{LibBi} modelling
language. At the end of Section \ref{sec:methods}, posterior distributions
are obtained for each model conditioned on simulated data sets, using
the methods presented. In Section \ref{sec:software}, the performance
of \pkg{LibBi} on various hardware platforms is compared. Section
\ref{sec:summary} summarises results. Section \ref{sec:supplementary}
provides information on available supplementary materials to reproduce
the example results throughout this work.

\section{State-space models}

\label{sec:models}

State-space models (SSMs) are suitable for modelling dynamical systems
that have been observed at one or more instances in time. They consist
of parameters $\boldsymbol{\Theta}\in\mathbb{R}^{N_{\theta}}$, a
latent continuous- or discrete-time state process $\mathbf{X}(t)\in\mathbb{R}^{N_{x}}$,
and an observed continuous- or discrete-time process $\mathbf{Y}(t)\in\mathbb{R}^{N_{y}}$.
A starting time is given by $t_{0}$, and an ordered sequence of observation
times by $t_{1},\ldots,t_{T}$.

The most general BHM over these random variables admits a joint density
of the form:
\begin{equation}
\underbrace{p\left(\mathbf{Y}(t_{1:T}),\mathbf{X}(t_{0:T}),\boldsymbol{\Theta}\right)}_{\textrm{joint}}=\underbrace{p\left(\mathbf{X}(t_{0:T}),\boldsymbol{\Theta}\right)}_{\textrm{prior}}\underbrace{p\left(\mathbf{Y}(t_{1:T})|\mathbf{X}(t_{0:T}),\boldsymbol{\Theta}\right)}_{\text{likelihood}}.\label{eqn:bhm-joint}
\end{equation}

\begin{figure}
\begin{centering}
\includegraphics[width=0.8\textwidth]{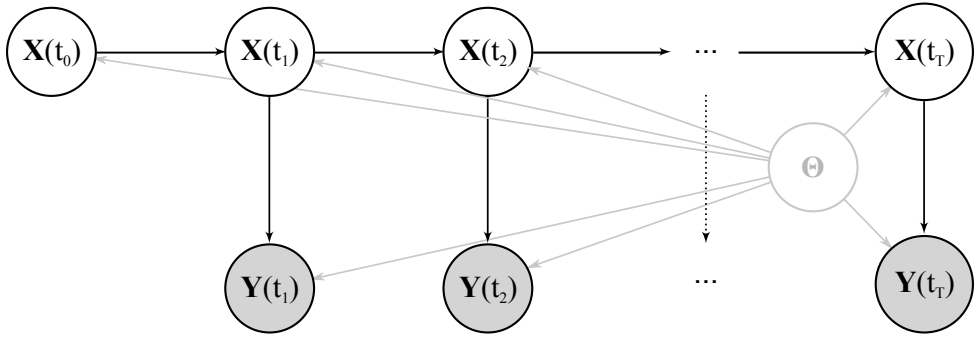}
\par\end{centering}

\caption{The state-space model, with parameters $\boldsymbol{\Theta}$, latent
Markov state process $\mathbf{X}(t_{0:T})$ and observation process
\textbf{$\mathbf{Y}(t_{1:T})$}.\label{fig:ssm}}
\end{figure}

The SSM class can be considered a specialisation of the general BHM
class. It imposes the Markov property on the state process $\mathbf{X}(t)$,
and at each time $t_{i}$ permits the observation $\mathbf{Y}(t_{i})$
to depend only on parameters $\boldsymbol{\Theta}$ and the state
$\mathbf{X}(t_{i})$. Formally, the joint density takes the form:

\begin{eqnarray}
\underbrace{p\left(\mathbf{Y}(t_{1:T}),\mathbf{X}(t_{0:T}),\boldsymbol{\Theta}\right)}_{\text{joint}} & = & \underbrace{\underbrace{p\left(\boldsymbol{\Theta}\right)}_{\text{parameter}}\underbrace{p\left(\mathbf{X}(t_{0})|\boldsymbol{\Theta}\right)}_{\text{initial}}\left(\prod_{i=1}^{T}\underbrace{p\left(\mathbf{X}(t_{i})|\mathbf{X}(t_{i-1}),\boldsymbol{\Theta}\right)}_{\textrm{transition}}\right)}_{\textrm{prior}}\underbrace{\left(\prod_{i=1}^{T}\underbrace{p\left(\mathbf{Y}(t_{i})|\mathbf{X}(t_{i}),\boldsymbol{\Theta}\right)}_{\text{observation}}\right)}_{\textrm{likelihood}},\label{eqn:ssm-joint}
\end{eqnarray}
also depicted as a graphical model in Figure \ref{fig:ssm}. The prior
density is factored into a parameter density, an initial state density,
and a product of one or more transition densities. The likelihood
function is given as a product of observation densities. Specific
models in the class may, of course, exploit further conditional independencies
and so introduce additional hierarchical structure. This is demonstrated
in the examples that follow.

\subsection{Examples}

Two example SSMs are introduced here. The first is a linear three-element
windkessel model of arterial blood pressure~\citep{Westerhof2009},
the second a nonlinear eight-dimensional Lorenz '96 model of chaotic
atmospheric processes~\citep{Lorenz2006}. In each case the original
deterministic model is introduced, then stochasticity added, and a
prior distribution over parameters specified, to massage it into the
SSM framework.

\subsubsection{Windkessel model}

\label{sec:windkessel-model}

A windkessel model can be used to relate blood pressure and blood
flow in the human arterial system~\citep{Westerhof2009}. The simplest
two-element windkessel~\citep{Frank1899} is physically inspired:
it couples a pump (the heart) and a chamber (the arterial system),
with fluid (blood) flowing from the pump to the chamber, and returning
via a closed loop. Air in the chamber is compressed according to the
volume of fluid (analogous to the compliance of arteries).

\begin{figure}
\begin{minipage}[t]{0.5\columnwidth}%
\includegraphics[width=1\linewidth]{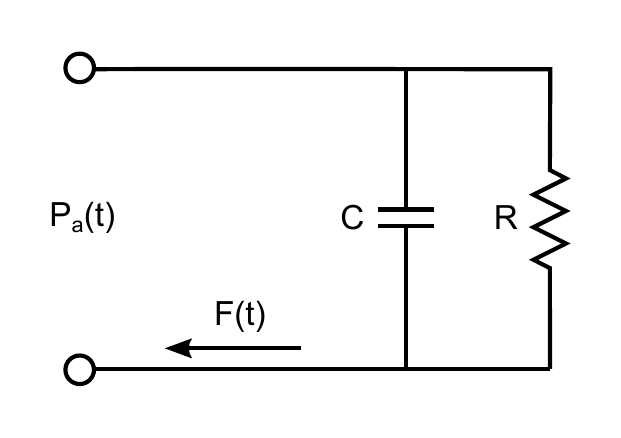}%
\end{minipage}\hfill{}%
\begin{minipage}[t]{0.5\columnwidth}%
\includegraphics[width=1\linewidth]{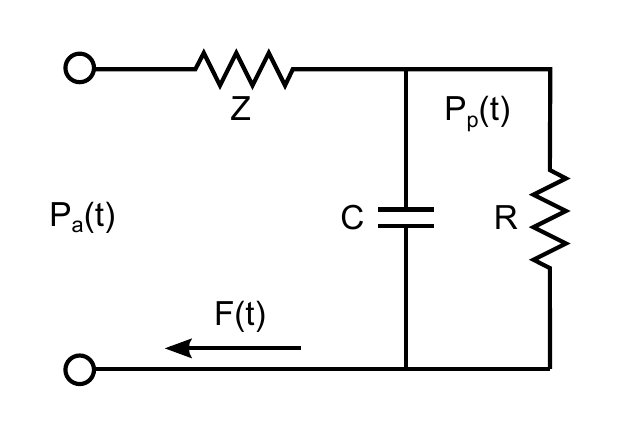}%
\end{minipage}

\caption{Windkessel models represented as circuit diagrams, \textbf{(left)}
two-element, and \textbf{(right)} three-element. $P_{a}(t)$ gives
aortal pressure and $P_{p}(t)$ peripheral pressure. $F(t)$ gives
blood flow.\label{fig:windkessel}}
\end{figure}

The two-element windkessel model is commonly represented as an RC
circuit as in Figure \ref{fig:windkessel} (left). Voltage models
blood pressure and current blood flow. A resistor models narrowing
vessel width in the periphery of the arterial system, and a capacitor
the compliance of arteries.

The equations of the two-element windkessel are readily derived from
circuit theory applied to Figure \ref{fig:windkessel} (left)~\citep{Kind2010}:
\[
\frac{1}{R}P_{p}(t)+C\frac{dP_{p}(t)}{dt}=F(t).
\]
This is a linear differential equation. Assuming a discrete time step
$\Delta t$ and constant flow $F$(t) over the interval $[t,t+\Delta t)$,
it may be solved analytically to give:
\begin{equation}
P_{p}(t+\Delta t)=\exp\left(-\frac{\Delta t}{RC}\right)P_{p}(t)+R\left(1-\exp\left(-\frac{\Delta t}{RC}\right)\right)F(t).\label{eqn:2-windkessel}
\end{equation}

The two-element windkessel captures blood pressure changes during
diastole (heart dilatation) well, but not during systole (heart contraction)~\citep{Westerhof2009}.
An improvement is the three-element windkessel in Figure \ref{fig:windkessel}
(right), which introduces additional impedence from the aortic valve.
Again using circuit theory, aortal pressure, $P_{a}$, may be related
to peripheral pressure, $P_{p}$, by~\citep{Kind2010}:
\begin{equation}
P_{a}(t)=P_{p}(t)+F(t)Z.\label{eqn:3-windkessel}
\end{equation}
The three-element windkessel is adopted henceforth. Blood flow, $F(t)$,
would usually be measured, but for demonstrative purposes it is simply
prescribed the following functional form:
\begin{equation}
F(t)=\begin{cases}
F_{\text{max}}\sin^{2}\left(\frac{\pi t}{T_{s}+T_{d}}\right) & \text{if }\mathrm{mod}(t,T_{s}+T_{d})\in[0,T_{s}]\\
0 & \text{if }\mathrm{mod}(t,T_{s}+T_{d})\in(T_{s},T_{s}+T_{d}),
\end{cases}\label{eqn:flow}
\end{equation}
where $F_{\text{max}}=500\,\text{ml s}^{-1}$ gives maximum flow,
$T_{s}=0.3\,\text{s}$ is time spent in systole, $T_{d}=0.5\,\text{s}$
is time spent in diastole, with $\mathrm{mod}(x,y)$ giving the non-integer
part of $x/y$. This models quickly increasing then decreasing blood
flow during systole, and no flow during diastole. The function is
discretised to time steps of $\Delta t$ and held constant in between.

An SSM can be constructed around the above equations. Input $F(t)$
is prescribed as above, and a Gaussian noise term $\xi(t)$ of zero
mean and variance $\sigma^{2}$ is introduced to extend the state
process from deterministic to stochastic behaviour (details below).
The SSM has parameters $\boldsymbol{\Theta\equiv}\left(R,C,Z,\sigma^{2}\right)^{T}$,
state $\mathbf{X}(t)\equiv P_{p}(t)$ and observation $\mathbf{Y}(t)\equiv P_{a}(t)$.
The time step is fixed to $\Delta t=10^{-2}$s.

The complete model is specified in the \pkg{LibBi} modelling language
in Figure \ref{fig:windkessel-bi}. The specification begins with
a \texttt{model} statement to name the model. It proceeds with the
time step size declared on line 5 as a constant value. Following this,
the four parameters of the model, \texttt{R} ($R$), \texttt{C} ($C$),
\texttt{Z} ($Z$) and \texttt{sigma2} ($\sigma^{2}$) are declared
on lines 7-10, the input \texttt{F} ($F$(t)) on line 11, the noise
term \texttt{xi} ($\xi(t)$) on line 12, the state variable \texttt{Pp}
($P_{p}(t)$) on line 13, and the observation \texttt{Pa} ($P_{a}(t)$)
on line 14.

\begin{figure}[p]
\parbox[t]{1\columnwidth}{%
\emph{Windkessel.bi}

\vspace{-1mm}
\rule[0.5ex]{1\columnwidth}{0.5pt}

\input{Windkessel.bi.tex}%
}

\caption{The windkessel SSM specified in the \pkg{LibBi} modelling language.\label{fig:windkessel-bi}}
\end{figure}

Recall (\ref{eqn:ssm-joint}), which gives the general joint distribution
of an SSM. The specific joint distribution for this SSM is:
\begin{eqnarray}
\underbrace{p(P_{a}(t_{1:T}),P_{p}(t_{0:T}),R,C,Z,\sigma^{2})}_{\textrm{joint}} & = & \underbrace{p(R)p(C)p(Z)p(\sigma^{2})}_{\textrm{parameter}}\underbrace{p(P_{p}(t_{0}))}_{\textrm{initial}}\nonumber \\
 &  & \hspace{-5cm}\times\left(\prod_{i=1}^{T}\underbrace{p(P_{p}(t_{i})|P_{p}(t_{i-1}),R,C,\sigma^{2},F(t_{i-1}))}_{\textrm{transition}}\right)\left(\prod_{i=1}^{T}\underbrace{p(P_{a}(t_{i})|P_{p}(t_{i}),Z,F(t_{i-1}))}_{\textrm{observation}}\right).\label{eqn:windkessel-joint}
\end{eqnarray}
Following the variable declarations in Figure \ref{fig:windkessel-bi}
are four \emph{blocks}, declared using the \texttt{sub} keyword, each
with the same name as, and describing the form of, one of the factors
in (\ref{eqn:windkessel-joint}). The prior distribution over parameters
is given by:
\begin{eqnarray*}
R & \sim & \textsc{Gamma}(2,0.9)\\
C & \sim & \textsc{Gamma}(2,1.5)\\
Z & \sim & \textsc{Gamma}(2,0.03)\\
\sigma^{2} & \sim & \textsc{Inv-Gamma}(2,25),
\end{eqnarray*}
described in the \texttt{parameter} block on lines 16-21 of Figure
\ref{fig:windkessel-bi}. The hyperparameters of these distributions
are guided by \citet{Kind2010}. The prior distribution over the initial
state is given by:
\begin{eqnarray*}
P_{p}(t_{0}) & \sim & \textsc{N}(90,15),
\end{eqnarray*}
where the second argument is the standard deviation. This is described
in the \texttt{initial} block on lines 23-25 of Figure \ref{fig:windkessel-bi}.
The transition model is a stochastic extension of (\ref{eqn:2-windkessel}).
This use of stochasticity might be interpreted as representing some
uncertainty in the formulation of the model given the biological phenomena
it is meant to represent. The noise term $\xi(t)$ is introduced additively
to blood flow, $F(t)$:
\[
P_{p}(t+\Delta t)=\exp\left(-\frac{\Delta t}{RC}\right)P_{p}(t)+R\left(1-\exp\left(-\frac{\Delta t}{RC}\right)\right)\left(F(t)+\xi(t)\right).
\]
This form is described in the \texttt{transition} block on lines 27-30
of Figure \ref{fig:windkessel-bi}. The observation model is based
on (\ref{eqn:3-windkessel}), with additional noise:
\[
P_{a}(t)\sim\textsc{N}(P_{p}(t)+ZF(t),2).
\]
This is described in the \texttt{observation} block on lines 32-34
of Figure \ref{fig:windkessel-bi}.

One additional block is specified for the windkessel model. This is
the \texttt{proposal\_parameter} block on lines 36-41. It gives the
proposal distribution over parameters that is used during marginal
Metropolis-Hastings sampling, described in Section \ref{sec:mcmc}.

Note that the input \texttt{F} does not appear on the left side in
any block. Instead, as an \texttt{input} variable, its value at each
time is drawn from an input file. This is prepared in advance from
(\ref{eqn:flow}) as a NetCDF file.

\subsubsection{Lorenz '96 model}

\label{sec:lorenz96-model}

\begin{figure}[tp]
\begin{centering}
\includegraphics[width=1\textwidth]{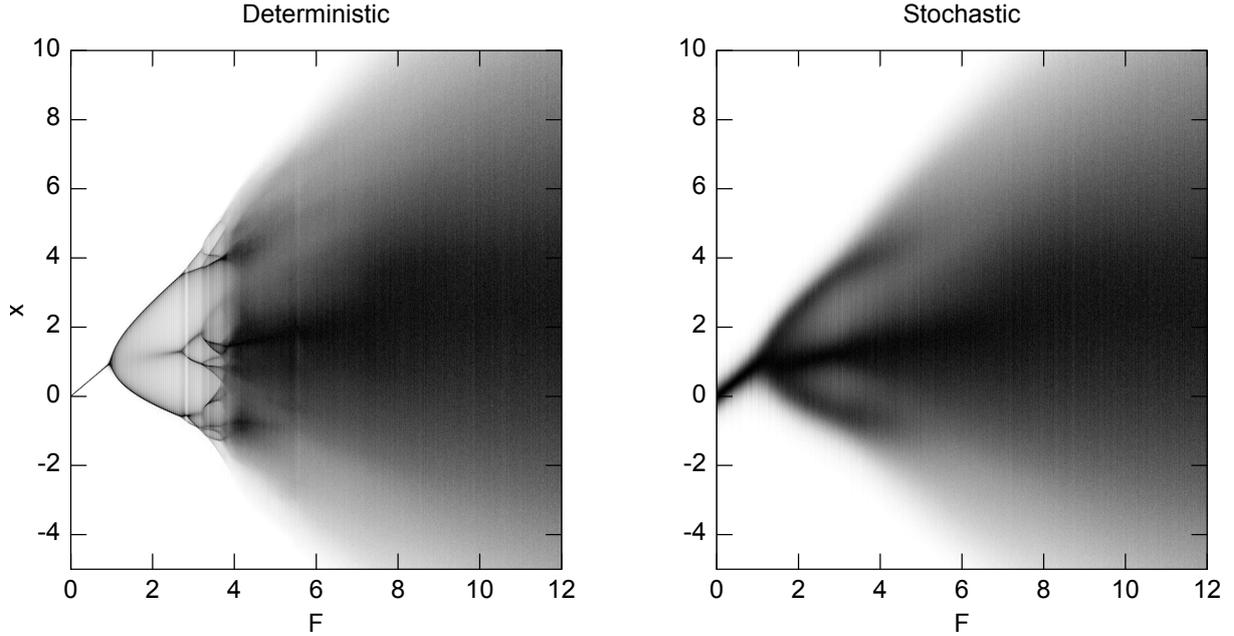}
\par\end{centering}

\caption{Density of the state variables $\mathbf{x}$ relative to the forcing
parameter $F$ in an eight-dimensional Lorenz '96 model. Densities
are rescaled by column, to reveal the bifurcation of \textbf{(left)}
the original deterministic model and \textbf{(right)} the modified
stochastic model with state marginalised over $\sigma^{2}$. Note
the way that decay behaviour at $0\leq F\leq1$ yields to periodic
behaviour for $F>1$, and eventually to chaotic behaviour. The prescribed
prior distribution $F\sim\mathrm{U}(8,12)$ places the state-space
model firmly within the chaotic regime.\label{fig:bifurc}}
\end{figure}

Lorenz '96 models are useful for the simulation of some important
atmospheric processes~\citep{Lorenz2006}. They can be challenging
to handle, owing to chaotic behaviour in most regimes. An $N$-dimensional
Lorenz '96 model over the vector $\mathbf{x}\in\mathbb{R}^{N}$ is
given by the ordinary differential equations (ODEs):
\begin{equation}
\frac{dx_{n}}{dt}=x_{n-1}(x_{n+1}-x_{n-2})-x_{n}+F,\label{eqn:lorenz96-ode}
\end{equation}
where indices into $\mathbf{x}$ are taken to be cyclic, so that $x_{n-N}\equiv x_{n}\equiv x_{n+N}$.
$F$ acts as a constant forcing term that induces various behaviours
ranging from decay, to periodicity, to chaos. The bifurcation diagram
of this deterministic system is given in Figure \ref{fig:bifurc}
(left). A deterministic model such as this is inappropriate for the
SSM framework. To derive a suitable stochastic model, the ODEs of
(\ref{eqn:lorenz96-ode}) can be converted to stochastic differential
equations (SDEs): 
\begin{equation}
dX_{n}=\left(X_{n-1}(X_{n+1}-X_{n-2})-X_{n}+F\right)\, dt+\sigma\, dW_{n}.\label{eqn:lorenz96-sde}
\end{equation}
Each $dW_{n}$ represents an increment of a standard Wiener process~\citep[\S3.8.1]{Gardiner2004},
and $\sigma$ is a parameter used to scale these. As the diffusion
term is additive the SDEs may be interpreted equivalently~\citep[p157]{Kloeden1992}
in either the Itô~\citep[\S4.2.1]{Gardiner2004} or Stratonovich~\citep[\S4.2.3]{Gardiner2004}
sense. The bifurcation diagram of this stochastic system is given
in Figure \ref{fig:bifurc} (right).

There is no analytical solution to integrate either the ODEs or SDEs
to obtain a form for $X_{n}$; they must be integrated numerically.
A classic fourth-order Runge-Kutta with fixed step-size $\Delta t$
suffices for this purpose, and was used for the original deterministic
model \citep{Lorenz2006}. For the stochastic model, by formally adopting
a Stratonovich interpretation of the (\ref{eqn:lorenz96-sde}), the
SDEs can be converted back to ODEs of the form \citep{Wilkie2004}:
\[
\frac{dx_{n}}{dt}=x_{n-1}(x_{n+1}-x_{n-2})-x_{n}+F+\sigma\frac{\Delta W_{n}}{\Delta t}.
\]
where each noise term $\Delta W_{n}\sim\mathrm{N}(0,\sqrt{\Delta t})$
is an increment of the Wiener process over the time step of size $\Delta t$.
In this form the classic fourth-order Runge-Kutta can be used again.
This final form is that used for the transition model of the SSM.

The SSM consists of two parameters, $\boldsymbol{\Theta}\equiv(F,\sigma^{2})^{T}$,
along with a state vector $\mathbf{x}(t)$ and observation vector
$\mathbf{y}(t)$%
, both of length 8. The complete model is specified in the \pkg{LibBi}
modelling language in Figure \ref{fig:lorenz96-bi}.

\begin{figure}[p]
\parbox[t]{1\columnwidth}{%
\emph{Lorenz96.bi}

\vspace{-1mm}
\rule[0.5ex]{1\columnwidth}{0.5pt}

\input{Lorenz96.bi.tex}%
}

\caption{The Lorenz '96 SSM specified in the \foreignlanguage{english}{\pkg{LibBi}
modelling language.\label{fig:lorenz96-bi}}\selectlanguage{english}%
}
\end{figure}

In Figure \ref{fig:lorenz96-bi}, a dimension named \texttt{n}, of
size 8, is first declared on line 5. It is given a cyclic boundary
condition, recalling that, in (\ref{eqn:lorenz96-ode}), indices are
interpreted cyclically. The two parameters of the model, \texttt{F}
($F)$ and \texttt{sigma2} ($\sigma^{2}$) are then declared on lines
9-10, the state vector \texttt{x} ($\mathbf{x}(t)$) on line 11, the
noise vector \texttt{deltaW} ($\Delta\mathbf{W}(t)$) on line 12,
and the observation vector \texttt{y} ($\mathbf{y}(t)$) on line 13.
The square bracket syntax is used to extend the vector variables over
the previously declared dimension \texttt{n}. Note that while\texttt{
$\mathbf{x}(t)$}, $\Delta\mathbf{W}(t)$ and $\mathbf{y}(t)$ all
vary in time, there is no explicit declaration for the time dimension
in \pkg{LibBi}.

Recall (\ref{eqn:ssm-joint}), which gives the general joint distribution
of an SSM. The specific joint distribution for the Lorenz '96 SSM
is:
\begin{eqnarray}
\underbrace{p(\mathbf{y}(t_{1:T}),\mathbf{x}(t_{0:T}),F,\sigma^{2})}_{\textrm{joint}} & = & \underbrace{p(F)p(\sigma^{2})}_{\textrm{parameter}}\underbrace{p(\mathbf{x}(t_{0}))}_{\textrm{initial}}\label{eqn:lorenz96-prior}\\
 &  & \hspace{-1cm}\times\left(\prod_{i=1}^{T}\underbrace{p(\mathbf{x}(t_{i})|\mathbf{x}(t_{i-1}),F,\sigma^{2})}_{\textrm{transition}}\right)\left(\prod_{i=1}^{T}\underbrace{{\textstyle \prod}_{j=1}^{N}p(y_{j}(t_{i})|x_{j}(t_{i}))}_{\textrm{observation}}\right).
\end{eqnarray}
The prior distribution over parameters is given by:
\begin{eqnarray*}
F & \sim & \textsc{U}(8,12)\\
\sigma^{2} & \sim & \textsc{Inv-Gamma}(2,0.25),
\end{eqnarray*}
described in the \texttt{parameter} block on lines 15-18 of Figure
\ref{fig:lorenz96-bi}. The prior distribution over the initial state
of $\mathbf{x}$ is given by:
\begin{eqnarray*}
x_{n}(t_{0}) & \sim & \textsc{U}(-1,3),
\end{eqnarray*}
described in the \texttt{initial} block on lines 20-22 of Figure \ref{fig:lorenz96-bi}.

The final form of the ODEs derived above is specified in the \texttt{transition}
block on lines 24-29 of Figure \ref{fig:lorenz96-bi}. Notice the
use of indexing on line 27 to show the relationship between elements
of the \texttt{x} vector, without the use of a loop.

Each component of the vector $\mathbf{y}$ is an observation of the
corresponding component of the vector $\mathbf{x}$, with some additive
Gaussian noise. The observation model is given by:
\begin{eqnarray*}
y_{n} & \sim & \textsc{N}(x_{n},0.5),
\end{eqnarray*}
where the second argument is the standard deviation. This is described
in the \texttt{observation} block on lines 31-33 of Figure \ref{fig:lorenz96-bi}.

Two additional blocks are specified for the Lorenz '96 model. These
are the \texttt{proposal\_parameter} and \texttt{proposal\_initial}
blocks on lines 35-38 and 40-42, respectively. These specify the proposal
distributions used for Metropolis-Hastings sampling, detailed in Section
\ref{sec:mcmc}.

\section{Inference methods}

\label{sec:methods}

An SSM is constructed as the joint distribution $p\left(\mathbf{Y}(t_{1:T}),\mathbf{X}(t_{0:T}),\boldsymbol{\Theta}\right)$,
typically factorised as in (\ref{eqn:ssm-joint}). The task of Bayesian
inference is to condition this joint distribution on some particular
data set, $\mathbf{Y}(t_{1:T})=\mathbf{y}(t_{1:T})$, to obtain the
posterior distribution $p(\mathbf{X}(t_{0:T}),\boldsymbol{\Theta}|\mathbf{y}(t_{1:T}))$.

The posterior distribution may be written:
\begin{eqnarray}
\underbrace{p(\mathbf{X}(t_{0:T}),\boldsymbol{\Theta}|\mathbf{y}(t_{1:T}))}_{\textrm{posterior}} & = & p\left(\boldsymbol{\Theta}|\mathbf{y}(t_{1:T})\right)p\left(\mathbf{X}(t_{0:T})|\boldsymbol{\Theta},\mathbf{y}(t_{1:T})\right).\label{eqn:posterior}
\end{eqnarray}
Obtaining the first factor constitutes \emph{parameter estimation},
while obtaining the second factor, conditioning on some particular
$\boldsymbol{\Theta}=\boldsymbol{\theta}$ drawn from the first, constitutes
\emph{state estimation}. Methods for Bayesian inference will in limited
cases derive a closed form for the posterior distribution. In most
cases this is unachievable, however, and either an approximate closed
form is fit, or Monte Carlo sampling is performed.

Because an SSM is a member of the broader BHM class, any method for
inference over BHMs can be applied to SSMs also. Methods that do not
exploit the additional hierarchical structure of SSMs may be sub-optimal,
however. SSMs also exhibit other qualities that can erode the effectiveness
of generic methods for BHMs. Common occurrences are strong autocorrelations
in the Markov process relative to the observation frequency \citep[e.g. ][]{Murray2013a},
and strong correlations between the model parameters and state \citep[e.g. ][]{Newman2009}.
The Gibbs sampler~\citep{Geman1984}, a mainstay of inference for
BHMs, is known to mix slowly in such conditions~\citep{Papaspiliopoulos2007}.
It is also common for the Markov process, while it can be simulated,
to not yield a useable closed-form transition density \citep[e.g.][]{Beskos2006,Fearnhead2008,Golightly2008,Golightly2011,Murray2011b,Murray2013a}.
This precludes an analytically derived conditional distribution for
Gibbs sampling, or even computation of the acceptance ratio for Metropolis-Hastings-within-Gibbs.

Fortunately, specialised methods for Bayesian inference with SSMs
do exist. Methods based on \emph{sequential Monte Carlo} (SMC) are
the focus of this work, and of the \pkg{LibBi} software. An important
exception is the Kalman filter \citep{Kalman1960}, which is optimal
for a further specialisation of the SSM class: that of linear-Gaussian
models. An introduction to the Kalman filter is given here also.

We begin with state estimation in Section \ref{sec:state-estimation},
which will lead into parameter estimation in Section \ref{sec:parameter-estimation}.

\subsection{State estimation}

\label{sec:state-estimation}

Sampling the second factor of (\ref{eqn:posterior}), conditioned
on a particular parameter setting $\boldsymbol{\Theta}=\boldsymbol{\theta}$,
constitutes state estimation. This has been well-studied in the Bayesian
filtering literature. Two methods, the Kalman filter and the particle
filter, are introduced here.

\subsubsection{The Kalman filter}

\label{sec:kf}

The Kalman filter~\citep{Kalman1960} can be used in the special
case of SSMs where both the transition and observation densities are
linear and Gaussian, and the initial state model is Gaussian%
\footnote{A fixed starting point is also admitted as a degenerate case of a
Gaussian distribution%
}. It is optimal in such cases, producing the exact closed-form solution.
Models that fit this class can be expressed in the following form:
\begin{eqnarray}
\mathbf{X}(t_{0}) & \sim & \textsc{N}\left(\boldsymbol{\mu}(t_{0}),\mathbf{U}(t_{0})\right)\label{eqn:linear-model-initial}\\
\mathbf{X}(t_{i})|\mathbf{X}(t_{i-1}) & \sim & \textsc{N}\left(\mathbf{F}^{\top}(t_{i})\mathbf{X}(t_{i-1}),\mathbf{Q}(t_{i})\right)\text{ for }i=1,\ldots,T\label{eqn:linear-model-state}\\
\mathbf{Y}(t_{i})|\mathbf{X}(t_{i}) & \sim & \textsc{N}\left(\mathbf{G}^{\top}(t_{i})\mathbf{X}(t_{i}),\mathbf{R}(t_{i})\right)\text{ for }i=1,\ldots,T,\label{eqn:linear-model-obs}
\end{eqnarray}
where $\textsc{N}(\boldsymbol{\mu},\mathbf{U})$ denotes the normal
distribution with mean vector $\boldsymbol{\mu}$ and square-root
of the covariance matrix $\mathbf{U}$. It is understood that all
symbols may depend on $\boldsymbol{\theta}$. The Kalman filter is
usually initialised with the mean vector $\boldsymbol{\mu}(t_{0})$
and covariance matrix $\boldsymbol{\Sigma}(t_{0})$ of the initial
state $p(\mathbf{X}(t_{0})|\boldsymbol{\theta})$. Here the \emph{square-root}
Kalman filter is presented, where the covariance matrix is replaced
with its upper-triangular Cholesky square-root $\mathbf{U}(t_{0})$,
i.e. $\boldsymbol{\Sigma}(t_{0})=\mathbf{U}^{\top}(t_{0})\mathbf{U}(t_{0})$.
Using $\mathbf{U}(t_{0})$ has certain numerical and computational
advantages; indeed most computational manipulations of Gaussian distributions
ultimately use the Cholesky square-root and not $\boldsymbol{\Sigma}(t_{0})$
directly.

After initialisation, the Kalman filter iterates through observation
times with interleaved \emph{prediction} and \emph{correction} steps.
Pseudocode is given in Algorithm \ref{alg:kalman-filter}. The prediction
step computes $p(\mathbf{X}(t_{i-1:i})|\mathbf{y}(t_{1:i-1}),\boldsymbol{\theta})$,
which is Gaussian with mean
\[
\left(\begin{array}{c}
\boldsymbol{\mu}(t_{i-1})\\
\hat{\boldsymbol{\mu}}(t_{i})
\end{array}\right)
\]
and upper-triangular Cholesky factor of the covariance matrix
\[
\left(\begin{array}{cc}
\mathbf{U}(t_{i-1}) & \mathbf{C}(t_{i})\\
\mathbf{0} & \hat{\mathbf{U}}(t_{i})
\end{array}\right).
\]
The cross term $\mathbf{C}(t_{i})$ is used later. The correction
step first computes $p(\mathbf{X}(t_{i}),\mathbf{Y}(t_{i})|\mathbf{y}(t_{1:i-1}),\boldsymbol{\theta})$,
also Gaussian, with mean
\[
\left(\begin{array}{c}
\hat{\boldsymbol{\mu}}(t_{i})\\
\boldsymbol{\nu}(t_{i})
\end{array}\right)
\]
and upper-triangular Cholesky factor of the covariance matrix
\[
\left(\begin{array}{cc}
\hat{\mathbf{U}}(t_{i}) & \mathbf{D}(t_{i})\\
\mathbf{0} & \mathbf{V}(t_{i})
\end{array}\right).
\]
It then conditions this on $\mathbf{Y}(t_{i})=\mathbf{y}(t_{i})$,
in the usual fashion for a Gaussian distribution, to obtain $p(\mathbf{X}(t_{i})|\mathbf{y}(t_{1:i}),\boldsymbol{\theta})$,
Gaussian with mean $\boldsymbol{\mu}(t_{i})$ and Cholesky factor
$\mathbf{U}(t_{i})$.

\begin{algorithm}[tp]
\input{kalman-filter.tex}\caption{The Kalman filter for given $\boldsymbol{\theta}$. This is presented
over a time interval $[t_{r},t_{s}]$, returning the marginal likelihood
$l=p(\mathbf{y}(t_{0:s})|\boldsymbol{\theta})$ and a state sample
$\mathbf{x}'(t_{0:s})\sim p(\mathbf{X}(t_{0:s})|\boldsymbol{\theta},\mathbf{y}(t_{0:s}))$,
so that it may be called from Algorithms \ref{alg:pmmh} \& \ref{alg:smc2}
later. The usual, standalone algorithm sets $r=0$ and $s=T$.\label{alg:kalman-filter}}
\end{algorithm}

In the context of parameter estimation in Section \ref{sec:parameter-estimation},
two further outputs are required of the Kalman filter. The first is
to deliver the likelihood of $\boldsymbol{\theta}$, marginalised
over $\mathbf{X}(t_{0:T})$. The computation is given on line \ref{line:kf-likelihood}
of Algorithm \ref{alg:kalman-filter} ($|\cdot|$ denotes the matrix
determinant and $\|\cdot\|$ the Euclidean norm). The second requirement
is to deliver a single sample $\mathbf{x}'(t_{0:T})\sim p(\mathbf{X}(t_{0:T})|\boldsymbol{\theta},\mathbf{y}(t_{1:T}))$.
This is achieved with a backward pass through time at the end of Algorithm
\ref{alg:kalman-filter}, in a manner similar to the Rauch-Tung-Striebel
smoother~\citep{Rauch1965}.

A number of Kalman-type filters exist for nonlinear and non-Gaussian
models. These include the mixture-of-Gaussians~\citep{Alspach1972},
extended~\citep{Smith1962}, ensemble~\citep{Evensen1994} and unscented~\citep{Julier1997}
Kalman filters and accompanying smoothers~\citep{Rauch1965,Sarkka2008}.
However, for nonlinear and non-Gaussian cases these are approximate
methods that introduce bias. The \emph{particle filter}~\citep{Gordon1993,Doucet2001}
is an alternative that does not, and admits any SSM of the form (\ref{eqn:ssm-joint}).
The particle filter is the focus for the nonlinear case in this work.
The approximate Kalman-type filters are not treated.

\subsubsection{The particle filter}

\label{sec:pf}

\begin{figure}[tp]
\begin{centering}
\includegraphics[width=0.8\textwidth]{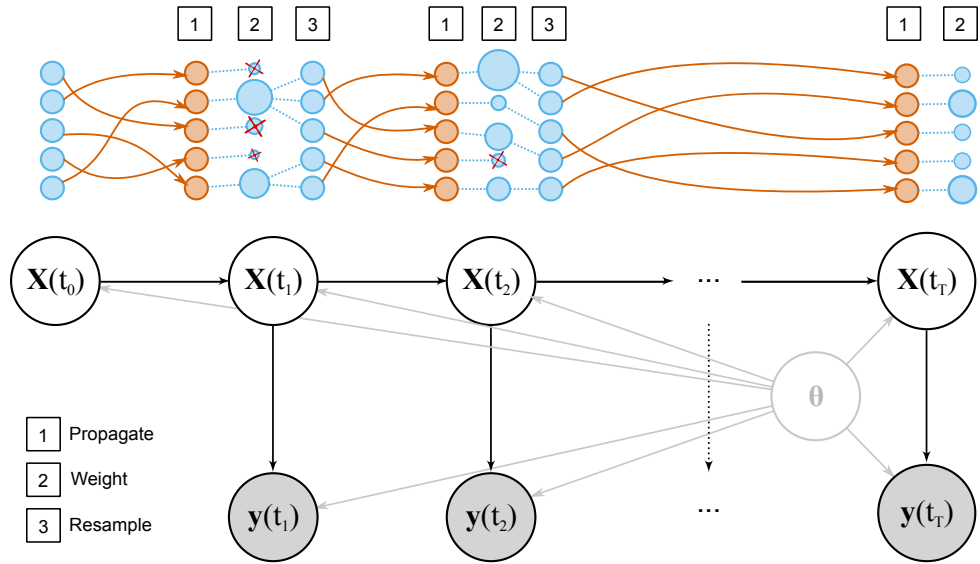}
\par\end{centering}

\caption{Visualisation of the particle filter algorithm where, at each time,
particles proceed through propagation, weighting and resampling steps.
Particles may be eliminated in each resampling step, with remaining
particles forming an ancestry tree.\label{fig:pf}}
\end{figure}

The particle filter~\citep{Gordon1993,Doucet2001}, of the family
of SMC methods, can be used for any SSM as defined by (\ref{eqn:ssm-joint}).
For this general form an analytical solution is not forthcoming, and
the particle filter instead relies on importance sampling. Numerous
variants are available, but the most basic, that of the \emph{bootstrap}
particle filter~\citep{Gordon1993}, is described in pseudocode in
Algorithm \ref{alg:particle-filter} and visualised in Figure \ref{fig:pf}.

\begin{algorithm}[tp]
\input{particle-filter.tex}\caption{The bootstrap particle filter for given $\boldsymbol{\theta}$. This
is presented over a time interval $[t_{r},t_{s}]$, returning an estimate
of the marginal likelihood $\hat{l}=p(\mathbf{y}(t_{0:s})|\boldsymbol{\theta})$
and an unbiased state sample $\hat{\mathbf{x}}'(t_{0:s})\sim p(\mathbf{X}(t_{0:s})|\boldsymbol{\theta},\mathbf{y}(t_{0:s}))$,
so that it may be called from Algorithms \ref{alg:pmmh} \& \ref{alg:smc2}
later. The usual, standalone algorithm sets $r=0$ and $s=T$.\label{alg:particle-filter}}
\end{algorithm}

For a given $\boldsymbol{\theta}$, the particle filter is initialised
by drawing $P_{x}$ number of random samples, $\mathbf{x}^{j}(t_{0})\sim p(\mathbf{X}(t_{0})|\boldsymbol{\theta})$,
for $j=1,\ldots,P_{x}$, and weighting each uniformly with $w^{j}(t_{0})=1/P_{x}$.
These are referred to as \emph{particles}. It proceeds sequentially
through observation times through a series of \emph{propagation},
\emph{weighting} and \emph{resampling} steps. In the propagation step
each particle is advanced to the next observation time with $\mathbf{x}^{j}(t_{i})\sim p(\mathbf{X}(t_{i})|\mathbf{x}^{a^{j}(t_{i})}(t_{i-1}),\boldsymbol{\theta})$,
where $a^{j}(t_{i})$ is the index of the particle's \emph{ancestor}
at the previous time, $t_{i-1}$ (more below). It is then weighted
with the likelihood of the new observation, $w^{j}(t_{i})=p(\mathbf{y}(t_{i})|\mathbf{x}^{j}(t_{i}),\boldsymbol{\theta})$.
The resampling step restores the particle stock to equal weights by
resampling particles with replacement, where the probability of each
particle being drawn is proportional to its weight $w^{j}$. Particles
with high weight tend to be replicated, while particles with low weight
tend to be eliminated. It is this process that determines the ancestor
indices for the next time propagation. 

The particle filter can be used to compute an unbiased estimate of
the likelihood of $\boldsymbol{\theta}$, marginalised over $\mathbf{X}(t_{0:T})$.
The computation~\citep{DelMoral2004} is given on line \ref{line:pf-likelihood}
of Algorithm \ref{alg:particle-filter}. It can also produce a sample
$\hat{\mathbf{x}}'(t_{0:T})\sim p(\mathbf{X}(t_{0:T})|\boldsymbol{\theta},\mathbf{y}(t_{1:T}))$
by tracing a single particle back through its ancestry~\citep{Andrieu2010}.
This occurs as the last step in Algorithm \ref{alg:particle-filter}.

\subsection{Parameter estimation}

\label{sec:parameter-estimation}

Sampling from the first factor of (\ref{eqn:posterior}) constitutes
parameter estimation, for which two approaches are given here. The
first is marginal Metropolis-Hastings~\citep{Metropolis1953,Hastings1970},
of the family of Markov chain Monte Carlo (MCMC), using either a Kalman
filter to compute the likelihood of parameters, marginalised over
the state, or a particle filter to estimate it. When a particle filter
is used the approach is more specifically known as particle marginal
Metropolis-Hastings (PMMH), from the family of particle Markov chain
Monte Carlo (PMCMC) methods~\citep{Andrieu2010}. The second approach
is sequential Monte Carlo (SMC)~\citep{DelMoral2006}. Recall that
the particle filter for state estimation is a type of SMC algorithm,
indeed it is prototypical. Similar methods may be employed for parameter
estimation, and this is considered here. Within the SMC algorithm
over parameters, a Kalman or particle filter is used to compute weights;
when the latter is used the method is more specifically known as SMC$^{2}$~\citep{Chopin2012}.

\subsubsection{Marginal Metropolis-Hastings}

\label{sec:mcmc}

The marginal Metropolis-Hastings algorithm is given in Algorithm \ref{alg:pmmh}.
It is initialised with some arbitrary setting of parameters $\boldsymbol{\theta}^{0}$,
then for $i=1,2,\ldots$, a new setting $\boldsymbol{\theta}'\sim q(\boldsymbol{\theta}'|\boldsymbol{\theta})$
is proposed from some proposal distribution $q$. The move is accepted
with probability given by the Metropolis-Hastings rule:
\begin{equation}
\min\left(1,\frac{p(\mathbf{y}(t_{1:T})|\boldsymbol{\theta}')p(\boldsymbol{\theta}')q(\boldsymbol{\theta}|\boldsymbol{\theta}')}{p(\mathbf{y}(t_{1:T})|\boldsymbol{\theta})p(\boldsymbol{\theta})q(\boldsymbol{\theta}'|\boldsymbol{\theta})}\right).\label{eqn:mh}
\end{equation}
If accepted then $\boldsymbol{\theta}^{i}=\boldsymbol{\theta}'$,
otherwise $\boldsymbol{\theta}^{i}=\boldsymbol{\theta}^{i-1}$. By
construction, the Markov chain $\boldsymbol{\theta}^{0},\boldsymbol{\theta}^{1},\ldots$
is ergodic to the posterior distribution, so that, after convergence,
states of the chain may be considered samples from it. The process
continues until as many samples as desired have been drawn, a number
denoted $P_{\theta}$.

Either a Kalman or particle filter may be used to compute or estimate
the marginal likelihood term $p(\mbox{\ensuremath{\mathbf{y}}}(t_{1:T})|\boldsymbol{\theta}')$
in (\ref{eqn:mh}) at each step. When estimated, a valid sampler is
still obtained~\citep{Andrieu2010}. The single state sample drawn
from the Kalman or particle filter completes the sample with a draw
from the second factor of (\ref{eqn:posterior}).

\begin{algorithm}[tp]
\input{pmmh.tex}

\caption{Marginal Metropolis-Hastings for parameter estimation. When the particle
filter is used (i.e. \proc{Particle-Filter} for \proc{Filter}),
the algorithm gives particle marginal Metropolis-Hastings (PMMH)~\citep{Andrieu2010}.\label{alg:pmmh}}
\end{algorithm}

\subsubsection{Sequential Monte Carlo}

\label{sec:smc}

An alternative approach to sampling the first factor of (\ref{eqn:posterior})
is to replace the MCMC over parameters with SMC over parameters. SMC
over parameters works similarly to SMC over state variables. It is
initialised by drawing $P_{\theta}$ number of random samples from
the prior distribution over parameters, $p(\boldsymbol{\Theta})$,
and weighting them uniformly with $v^{j}(t_{0})=1/P_{\theta}$ for
$j=1,\ldots,P_{\theta}$. These are referred to as \emph{$\theta$-particles}
and \emph{$\theta$-weights}. It proceeds sequentially through observation
times with a series of propagation, weighting and resampling steps,
just as for the particle filter for state estimation, along with a
new \emph{rejuvenation} step. Pseudocode for the algorithm is given
in Algorithm \ref{alg:smc2}.

\begin{algorithm}[tp]
\input{smc2.tex}

\caption{Sequential Monte Carlo (SMC) for parameter estimation. When the particle
filter is used (i.e. \proc{Particle-Filter} for \proc{Filter}),
the algorithm gives SMC$^{2}$~\citep{Chopin2012}.\label{alg:smc2}}
\end{algorithm}

To each $\theta$-particle is attached a Kalman or particle filter.
Propagating a $\theta$-particle involves advancing the attached filter
through time. Weighting of a $\theta$-particle uses the marginal
likelihood $p(\mathbf{y}(t_{i})|\mathbf{\boldsymbol{\theta}}^{j}(t_{i}))$,
computed by the attached Kalman filter or estimated by the attached
particle filter. Weighting with an unbiased estimate of that marginal
likelihood gives a valid SMC algorithm as in the random-weight particle
filter~\citep{Fearnhead2008,Fearnhead2010a}, and is more specifically
called SMC$^{2}$~\citep{Chopin2012}. Resampling involves drawing
a new set of unweighted $\theta$-particles by weight, along with
their attached filters.

The resampling of $\theta$-particles at each time $t_{i}$ depletes
the number of unique values represented. Resampling has the same effect
in the particle filter for state estimation, but in that case particles
diversify again in the next propagation step. As parameters do not
change in time, they cannot diversify in this way. To fix this, an
additional step, called the \emph{rejuvenation} step, is inserted
after the resampling of $\theta$-particles~\citep{Chopin2012}.
The aim of the step is to diversify the values of $\theta$-particles
while preserving their distribution. To this end it is sufficient
to take a single marginal Metropolis-Hastings step for each $\theta$-particle,
$\boldsymbol{\mathbf{\theta}}^{j}(t_{i})$. This works by proposing
a move to a new value $\boldsymbol{\mathbf{\theta}}'(t_{i})\sim q(\boldsymbol{\mathbf{\theta}}'(t_{i})|\boldsymbol{\mathbf{\theta}}^{j}(t_{i}))$,
estimating the marginal likelihood $\hat{p}(\mathbf{y}(t_{1:i})|\boldsymbol{\mathbf{\theta}}'(t_{i}))$
with the Kalman or particle filter, and then accepting or rejecting
the move using the acceptance probability (\ref{eqn:mh}). Line 7
of Algorithm \ref{alg:smc2} achieves this by calling the \proc{Marginal-Metropolis-Hastings}
function of Algorithm \ref{alg:pmmh}.

\subsection{Parallelisation}

The methods presented exhibit varying degrees of parallelisability,
and therefore suitability to modern computing hardware. SMC is particularly
promising in this regard~\citep{Lee2010}. The propagation, weighting
and (in the case of parameter estimation) rejuvenation steps can be
performed in parallel for each ($\theta$-)particle. There is limited
scope for parallelisation of the Kalman filter: its matrix operations
can be multithreaded or even performed on GPU, but this will only
be faster for very large matrices~\citep{Song2012}, and may be slower
for small matrices.

For the particle filter, the minimum degree of parallelism is $P_{x}$,
the number of particles, with the potential for further parallelisation
according to model-specific structure. The bottleneck in high-performance
implementation of the particle filter is the resampling step. Resampling
algorithms, such as the multinomial, systematic or stratified~\citep{Kitagawa1996}
approaches, require a collective operation over the weight vector.
This means that all threads must synchronize. The development of asynchronous
resampling methods to alleviate this bottleneck is still an active
area of research~\citep{Murray2011a,Murray2013,DelMoral2013}. In
the meantime, the particle filter is best suited to shared-memory
architectures where the collective operation can be performed efficiently,
although approaches for distributed-memory resampling have been proposed~\citep{Bolic2005}.

Marginal Metropolis-Hastings is itself a sequential method, but inherits
the degree of parallelism of the filter used at each step. Again,
the Kalman filter offers limited scope, but using the particle filter
(the PMMH sampler) gives $P_{x}$-way parallelism. The setting of
$P_{x}$ is complicated in this context, however. The tradeoff is
to maximise the mixing rate of the Metropolis-Hastings chain against
the computational expense of running $P_{x}$ particles. The optimal
choice relates to the variance in the likelihood estimator \citep{Doucet2013,Murray2013a}.
Increasing $P_{x}$ decreases this variance, but does not necessarily
improve the real-time mixing of the Metropolis-Hastings chain for
a fixed computational budget. Because arbitrarily increasing $P_{x}$
to consume all available hardware resources has depreciating returns
on mixing rate, there are limits on the degree of parallelism of PMMH.

A possible solution is to run multiple marginal Metropolis-Hastings
chains. This also has limits. In practice, one usually removes some
number, $B_{\theta}$, of steps from the start of each chain to correct
for the initialisation bias of $\boldsymbol{\theta}^{0}$. $B_{\theta}$
depends on the autocorrelation of the chain, and should be sufficiently
large for the influence of $\boldsymbol{\theta}^{0}$ to be forgotten.
With multiple chains, each chain must take at least $B_{\theta}$
steps before drawing one or more samples that will be preserved. This
imposes a serial limitation on the maximum speedup achievable by parallelisation
with multiple chains, which may be quantified by Amdahl's law~\citep{Amdahl1967}
as $P_{\theta}/B_{\theta}$. The performance gains of multiple chains
might therefore be disappointing without some additional strategy
to reduce $B_{\theta}$. Adaptation~\citep{Gilks1998,Haario2001,Andrieu2008}
and tempering~\citep{Frantz1990,Geyer1991,Marinari1992,Geyer1995,Neal1996}
are both established means of reducing $B_{\theta}$ for single chains.
Using these for each chain in isolation can reduce $B_{\theta}$,
but only by a factor that is independent of the number of chains.
Reducing $B_{\theta}$ relative to the number of chains is to be preferred.
Population MCMC~\citep{Laskey2003} attempts this via an evolutionary~\citep{Back1996}
selection, crossing and mutation of multiple chains. \citet{Craiu2009}
more explicitly target the posterior with an ensemble of chains, using
the covariance of samples across all chains to adapt the proposal
covariance of individual chains. This remains an active area of research.

Using SMC for parameter estimation gives at least a $P_{\theta}$-way
parallelism, even with the Kalman filter. Using the particle filter
(the SMC$^{2}$ method) has a much higher degree of parallelism, at
$P_{\theta}P_{x}$. This is very promising. As for the particle filter,
however, the resampling step is synchronous, which can be a bottleneck
to the scaling of the algorithm. This also remains an active area
of research.

\subsection{Examples}

The example models introduced in Section \ref{sec:models} are used
within \pkg{LibBi} to demonstrate the inference methods above. The
three-element windkessel model is an example where Kalman filter-based
methods are suitable, while the nonlinear Lorenz '96 model requires
the particle filter. In both cases simulated data sets are used.

\pkg{LibBi} provides a \texttt{libbi} command to access its functionality
from the command line. It is used as follows:

\begin{lstlisting}
libbi command [options...]
\end{lstlisting}
where \texttt{command} is the particular command to execute, followed
by zero or more command-line options to specify input files and configuration
parameters. The pertinent command for these examples is \texttt{sample},
which draws samples from the joint, prior or posterior distribution.
Options may be given on the command line itself, or, as is often convenient,
by listing them in a configuration file and giving the name of that
file on the command line instead, like so:

\begin{lstlisting}
libbi sample @config.conf
\end{lstlisting}
All of the examples use configuration files in this way. The contents
of these files are given in Figure \ref{fig:windkessel-conf}.

\subsubsection{Windkessel example}

\label{sec:windkessel-methods}

\begin{figure}
\parbox[t]{1\columnwidth}{%
\emph{prior.conf}

\vspace{-1mm}
\rule[0.5ex]{1\columnwidth}{0.5pt}

\vspace{-4mm}
\begin{verbatim}
--target prior
--model-file Windkessel.bi
--nsamples 20000
--start-time 0.0
--end-time 2.4
--noutputs 240
--input-file data/input.nc
--output-file results/prior.nc
\end{verbatim}%
}

\parbox[t]{1\columnwidth}{%
\emph{posterior.conf}

\vspace{-1mm}
\rule[0.5ex]{1\columnwidth}{0.5pt}

\vspace{-4mm}
\begin{verbatim}
--target posterior
--model-file Windkessel.bi
--filter kalman
--nsamples 20000
--start-time 0.0
--end-time 2.4
--noutputs 240
--input-file data/input.nc
--obs-file data/obs.nc
--output-file results/posterior.nc
\end{verbatim}%
}

\caption{\pkg{LibBi} configuration files, containing command-line options,
for the the windkessel example.\label{fig:windkessel-conf}}
\end{figure}

The three-element windkessel model has Gaussian initial state, linear
and Gaussian transition model, and linear and Gaussian observation
model. Kalman filter-based methods are suitable in this case. The
model can be coerced into the matrix form of (\ref{eqn:linear-model-initial}-\ref{eqn:linear-model-obs}),
but it is not necessary to do so by hand: \pkg{LibBi} uses symbolic
differentiation to derive the matrix form internally. Consequently,
no changes are required to the model, as specified in Appendix \ref{app:windkessel-model},
to apply the Kalman filter.

\begin{figure}[p]
\centering{}\includegraphics[width=0.9\textwidth]{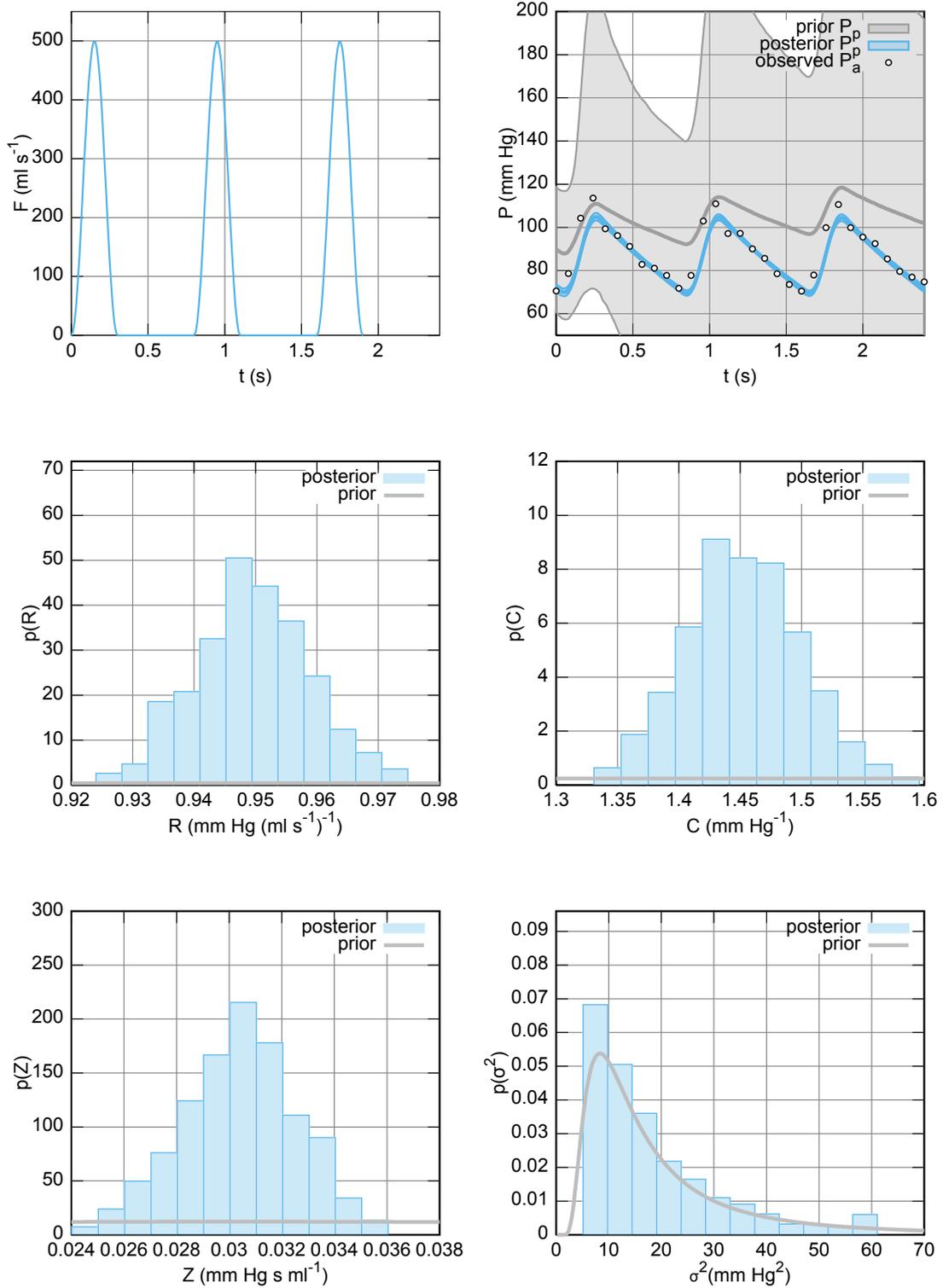}\caption{\textbf{(top left)} input blood flow $F(t)$, \textbf{(top right)}
prior and posterior distribution of the state $P_{p}(t)$, bold lines
giving the median and shaded regions the 95\% credibility interval
at each time, with observations $P_{a}(t)$ plotted over the top,
and \textbf{(remainder)} histograms of the posterior samples of parameters
against prior distributions. Note that the 95\% credibility interval
for the state estimate of $P_{p}(t)$ in the top right is difficult
to distinguish owing to its narrowness. \label{fig:windkessel-results}}
\end{figure}

The windkessel model requires two input files: one giving the values
of the input variable $F(t)$, and one giving the values of the observed
variable $P_{a}(t)$. The input is shown in the top-left plot of Figure
\ref{fig:windkessel-results}, and observations in the top-right.
These have been prepared in advance as NetCDF files \texttt{data/input.nc}
and \texttt{data/obs.nc}, available in the supplementary materials
(see Section \ref{sec:supplementary}).

A first step is often to simulate the model's prior distribution.
This is useful to validate the model specification. Command-line options
for this purpose are given in the \texttt{prior.conf} file in Figure
\ref{fig:windkessel-conf}. The command is then:

\begin{lstlisting}
libbi sample @prior.conf
\end{lstlisting}
with results output to the NetCDF file \texttt{results/prior.nc} and
shown in the top-right of Figure \ref{fig:windkessel-results}. The
posterior distribution can be sampled using the options in the \texttt{posterior.conf}
file in Figure \ref{fig:windkessel-conf}. Note in particular the
\texttt{-{}-filter kalman} option to indicate that a Kalman filter
should be used for state estimates. The default sampling method is
marginal Metropolis-Hastings, so no options are required to select
this. The command is:

\begin{lstlisting}
libbi sample @posterior.conf
\end{lstlisting}
with results output to the NetCDF file \texttt{results/posterior.nc}.
State estimates are shown in the top-right of Figure \ref{fig:windkessel-results},
and parameter estimates in the lower four plots.

\subsubsection{Lorenz '96 example}

\label{sec:lorenz96-methods}

\begin{figure}
\parbox[t]{1\columnwidth}{%
\emph{prior.conf}

\vspace{-1mm}
\rule[0.5ex]{1\columnwidth}{0.5pt}

\vspace{-4mm}
\begin{verbatim}
--target prior
--model-file Lorenz96.bi
--end-time 3
--noutputs 60
--nsamples 100000
--output-file results/prior.nc
\end{verbatim}%
}

\parbox[t]{1\columnwidth}{%
\emph{posterior.conf}

\vspace{-1mm}
\rule[0.5ex]{1\columnwidth}{0.5pt}

\vspace{-4mm}
\begin{verbatim}
--target posterior
--model-file Lorenz96.bi
--end-time 2
--noutputs 40
--nsamples 100000
--nparticles 512
--obs-file data/obs_sparse.nc
--output-file results/posterior.nc
--with-transform-initial-to-param
\end{verbatim}%
}

\parbox[t]{1\columnwidth}{%
\emph{prediction.conf}

\vspace{-1mm}
\rule[0.5ex]{1\columnwidth}{0.5pt}

\vspace{-4mm}
\begin{verbatim}
--target prediction
--model-file Lorenz96.bi
--start-time 2
--end-time 3
--noutputs 20
--nsamples 100000
--init-file results/posterior.nc
--output-file results/prediction.nc
\end{verbatim}%
}

\caption{\pkg{LibBi} configuration files, containing command-line options,
for the the Lorenz '96 example.\label{fig:lorenz96-conf}}
\end{figure}

A similar procedure is followed for the Lorenz '96 model, although
a prediction forward in time will also be made. Because it is a nonlinear
model, the particle filter is used for state estimation and marginal
likelihood estimates.

The Lorenz '96 model requires one input file, containing the observations.
This has been prepared in advance as \texttt{data/obs\_sparse.nc},
available in the supplementary materials (see Section \ref{sec:supplementary}).
It contains observations of the first four components of $\mathbf{y}(t)$
at every other time step (recall $\Delta t=0.05$) on the interval
$t=[0,3]$. It is an example of how sparse or missing data can be
handled by \pkg{LibBi}. The aim is to condition the joint distribution
on those observations that fall in the interval $t=[0,2]$, and set
aside the remainder to validate a forward prediction on the interval
$t=(2,3]$.

\begin{figure}
\includegraphics[width=1\textwidth]{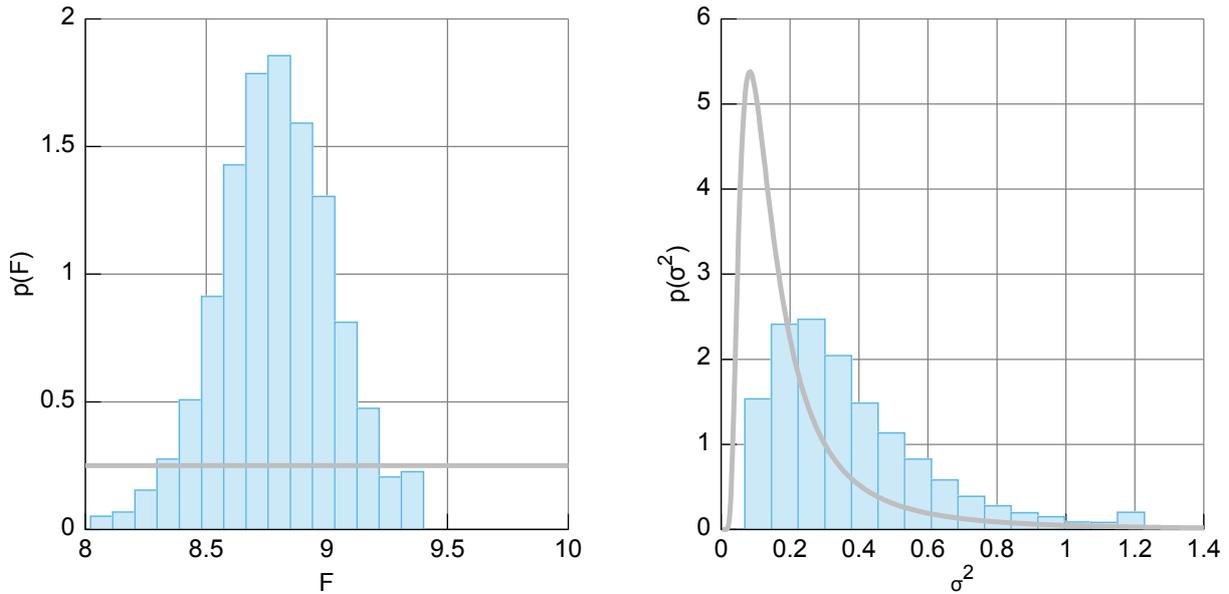}\caption{Parameter estimates for the Lorenz '96 case study, \textbf{(left)}
$F$ and \textbf{(right)} $\sigma^{2}$. In both cases the blue histogram
summarises posterior samples, while the grey curve denotes the prior
distribution.\label{fig:param}}
\end{figure}

\begin{figure}
\includegraphics[width=1\textwidth]{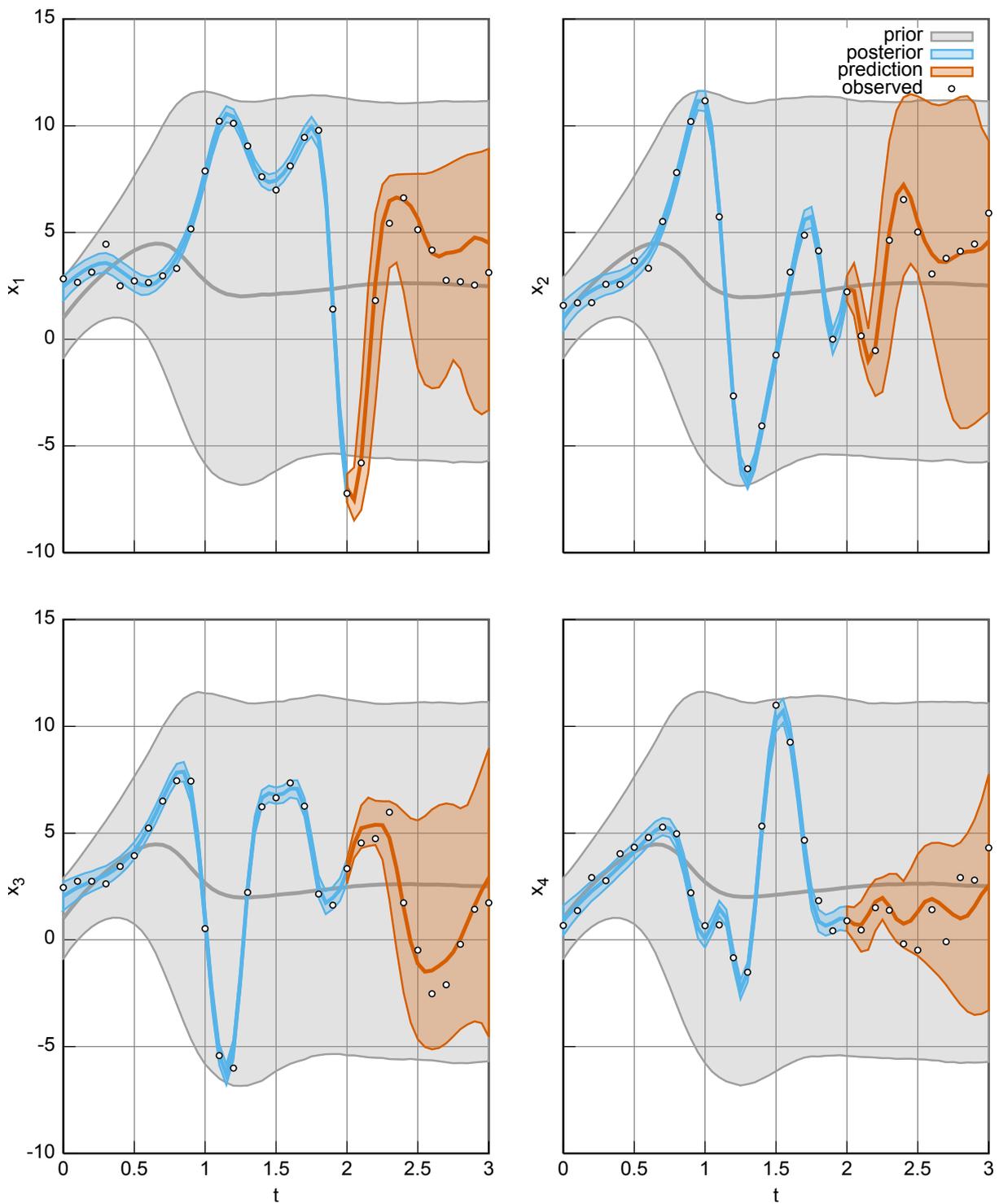}\caption{State estimates for the Lorenz '96 case study, giving the four components
of the state vector $\mathbf{x}(t)$ that are observed with additional
noise. Grey denotes the prior distribution, blue the posterior and
red the prediction. For each, the central line gives the time-marginal
median and the shaded region about it the 95\% credibility interval.
White circles indicate the observations. Observations that fall in
the time interval of prediction are witheld from the inference, but
shown for validation purposes.\label{fig:state1}}
\end{figure}

\begin{figure}
\includegraphics[width=1\textwidth]{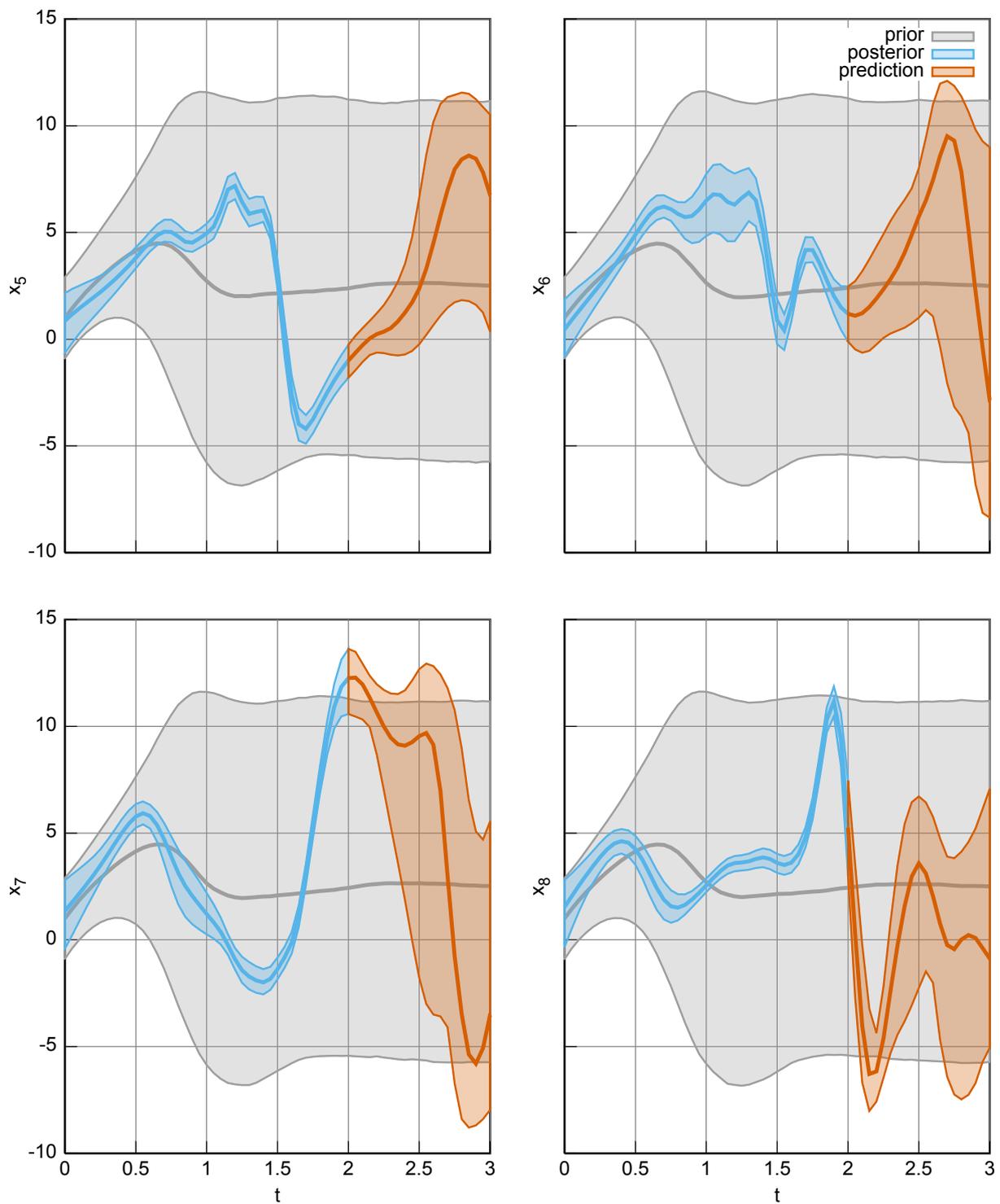}\caption{State estimates for the Lorenz '96 case study, giving the four components
of the state vector $\mathbf{x}(t)$ that are not observed. See Figure
\ref{fig:state1} for explanation of components.\label{fig:state2}}
\end{figure}

The first task is to simulate the prior distribution. The \texttt{prior.conf}
configuration file in Figure \ref{fig:lorenz96-conf} is set up for
this purpose. The command:

\begin{lstlisting}
libbi sample @prior.conf
\end{lstlisting}
outputs samples to the \texttt{results/prior.nc} file. These are shown
in grey in Figures \ref{fig:state1} and \ref{fig:state2}. The posterior
is sampled with options from the \texttt{posterior.conf} configuration
file of Figure \ref{fig:lorenz96-conf}. The particle filter is the
default option for state estimation, and marginal Metropolis-Hastings
for parameter estimation, so no options to select an appropriate method
are required. The command is:

\begin{lstlisting}
libbi sample @posterior.conf
\end{lstlisting}
Results are output to \texttt{results/posterior.nc}, and plot in blue
in Figures \ref{fig:param}-\ref{fig:state2}. SMC$^{2}$ does work
to sample from the posterior distribution also. It may be selected
by adding the \texttt{-{}-sampler smc2} option (see the \texttt{posterior\_smc2.conf}
file in the supplementary materials).

Finally, a prediction can be made. The configuration file \texttt{prediction.conf}
of Figure \ref{fig:lorenz96-conf} is set up for this. The output
file of the posterior sample is used as an input file for the prediction
(\texttt{-{}-init-file results/posterior.nc}) so as to extend the
state estimate forward in time to form a posterior prediction. The
command is:

\begin{lstlisting}
libbi sample @prediction.conf
\end{lstlisting}
Results are output to \texttt{results/prediction.nc}, and plot in
red in Figures \ref{fig:state1}-\ref{fig:state2}. Note that the
observations on $t=(2,3]$ do not factor in to the prediction, they
are shown in Figure \ref{fig:state1} only for validation of the prediction.

\section[The LibBi software]{The \pkg{LibBi} software}

\label{sec:software}

\pkg{LibBi} is made up of a \proglang{C++} template library and
\proglang{Perl} frontend. The \proglang{C++} template library provides
the inference methods, along with supporting functionality for, among
other things, input and output, memory management, matrices and vectors,
and pseudorandom number generation. The \proglang{Perl} frontend
parses the \pkg{LibBi} modelling language, generates model-specific
\proglang{C++} code using the Perl Template Toolkit%
\footnote{\url{http://www.template-toolkit.org}%
}, uses a GNU Autotools%
\footnote{\url{http://www.gnu.org/software/autoconf/} and \url{http://www.gnu.org/software/automake/}%
} build system to compile and link this against the \proglang{C++}
template library, and finally runs the program. Code is configured
for the hardware platform according to options set by the user when
calling the \texttt{libbi} command. \pkg{LibBi} supports several
hardware architectures and high-performance computing technologies,
including vector SIMD (Single Instruction Multiple Data) operations
on CPU using SSE (Streaming SIMD Extensions), multithreading on multicore
CPUs using OpenMP%
\footnote{\url{http://www.openmp.org}%
}, general-purpose GPU programming on NVIDIA GPUs using CUDA (Compute
Unified Device Architecture)%
\footnote{\url{http://www.nvidia.com/cuda/}%
}, and distributed-memory computing using MPI (Message Passing Interface)%
\footnote{\url{http://www.mpi-forum.org}%
}. The compilation process is hidden from the user. Intermediate files
are preserved in a hidden directory to avoid repeated effort, although
a short wait is noticeable after making changes to a model that require
recompilation.

The \pkg{LibBi} modelling language is described with an LALR grammar~\citep{DeRemer1969}.
Tools such as Yacc%
\footnote{\url{http://dinosaur.compilertools.net/yacc/}%
} and GNU Bison%
\footnote{\url{http://www.gnu.org/software/bison/}%
} may be used to compile parsers from the grammar. Yapp%
\footnote{\url{http://search.cpan.org/~fdesar/Parse-Yapp-1.05/yapp/}%
}, a Yacc-like parser compiler for \proglang{Perl}, is used for this
purpose. Input and output files use the standard NetCDF%
\footnote{\url{http://www.unidata.ucar.edu/software/netcdf/}%
} format, based on HDF5%
\footnote{\url{http://www.hdfgroup.org/HDF5/}%
}, and are readily created and analysed within various mathematical
and statistical packages such as MATLAB%
\footnote{\url{http://www.mathworks.com/matlab/}%
}, R%
\footnote{\url{http://www.R-project.org}%
}~\citep{R} and GNU Octave%
\footnote{\url{http://www.octave.org}%
}~\citep{Eaton2002}. The workflow for \pkg{LibBi} typically sees
the user prepare input files in their preferred statistical package,
run \pkg{LibBi} from the command-line, then return to their statistical
package to summarise and visualise the results. It is hoped that closer
integration will be realised in future.

Existing software for general BHMs exists, including the BUGS pair
\pkg{WinBUGS}%
\footnote{\url{http://www.mrc-bsu.cam.ac.uk/bugs/}%
}~\citep{Lunn2000} and more recently \pkg{OpenBUGS}%
\footnote{\url{http://www.openbugs.info}%
}~\citep{Lunn2012}, as well as \pkg{JAGS}%
\footnote{\url{http://mcmc-jags.sourceforge.net}%
}~\citep{Plummer2003} and \pkg{Stan}%
\footnote{\url{http://mc-stan.org}%
}~\citep{STAN}. Specialist software for SSMs also exists, notably
\pkg{BiiPS}%
\footnote{\url{https://alea.bordeaux.inria.fr/biips/}%
}. It is against these programs that \pkg{LibBi} might be most closely
compared. All of these existing packages accept models specified in
the \proglang{BUGS} language or extensions of it~\citep{Plummer2003},
while \pkg{LibBi} prescribes its own, albeit similar, language. \pkg{WinBUGS},
\pkg{OpenBUGS}, \pkg{JAGS} and \pkg{BiiPS} build a data structure
from the model which is manipulated internally by the client program,
while \pkg{Stan} and \pkg{LibBi} take a code generation approach.
\pkg{WinBUGS}, \pkg{OpenBUGS} and \pkg{JAGS} target general BHMs,
while \pkg{BiiPS} and \pkg{LibBi} target the more-specific class
of SSMs. In line with this more specialised focus, \pkg{BiiPS} and
\pkg{LibBi} use SMC as a staple method, rather than Gibbs (as in
\pkg{WinBUGS}, \pkg{OpenBUGS} and \pkg{JAGS}) or Hamiltonian Monte
Carlo~\citep{HoffmanInpress} (as in \pkg{Stan}). The most notable
distinction of \pkg{LibBi} against all of these packages is its high-performance
computing focus, where its hardware support, particularly for GPUs
and distributed clusters, is unique.

The \proglang{C++} template library component of \pkg{LibBi} might
also be compared to similar libraries such as \pkg{SMCTC}%
\footnote{\url{http://www2.warwick.ac.uk/fac/sci/statistics/staff/academic-research/johansen/smctc/}%
}~\citep{Johansen2009}. The template metaprogramming techniques advocated
in \citet{Johansen2009} are mirrored in \pkg{LibBi}. But where a
library requires its user to program their model to a prescribed interface,
an additional code generator component, as in \pkg{LibBi}, automates
compliance with the interface from a higher-level language in which
the model is specified. In \pkg{LibBi} this also facilitates the
generation of code for different combinations of methods and hardware.

\subsection{Performance results}

\label{sec:performance}

This section offers a comparison of \pkg{LibBi} performance under
different hardware configurations. The Lorenz '96 model is used for
this purpose. Experiments are conducted on a single machine with two
8-core Intel Xeon E5-2650 CPUs, three NVIDIA Tesla 2075 GPUs, and
128 GB main memory. The salient technical specifications of these
devices are given in Table \ref{tab:devices}. The particle filter,
PMMH and SMC$^{2}$ methods described in Section \ref{sec:methods}
are configured according to Table \ref{tab:method-config} using hardware
configurations in Table \ref{tab:system-config}. All methods are
tested with the first six configurations in Table \ref{tab:system-config}.
The SMC$^{2}$ method is also tested with the last two configurations,
which use MPI to distribute $\boldsymbol{\theta}$-particles across
multiple processes.%

\begin{table}[tp]
\centering{}\input{hardware.tex}\caption{Capabilities of the CPU and GPU devices used for empirical experiments,
from a programming perspective. Specifications are obtained from \texttt{/proc/cpuinfo}
for the CPU, and the CUDA SDK \texttt{deviceQuery} program for the
GPU.\label{tab:devices}}
\end{table}

\begin{table}[tp]
\centering{}%
\begin{tabular}{rcccc}
 & $P_{\theta}$ & $P_{x}$ & $T$ & $t_{T}$\tabularnewline
 & \texttt{-{}-nsamples }\texttt{\emph{n}} & \texttt{-{}-nparticles }\texttt{\emph{n}} & \texttt{-{}-noutputs }\texttt{\emph{n}} & \texttt{-{}-end-time }\texttt{\emph{n}}\tabularnewline
\hline 
Particle filter & 1 & 8192 & 40 & 2\tabularnewline
PMCMC & 500 & 8192 & 40 & 2\tabularnewline
SMC$^{2}$ & 192 & 8192 & 40 & 2\tabularnewline
\hline 
\end{tabular}\caption{Method configurations for timing results.\label{tab:method-config}}
\end{table}

\begin{table}[tp]
\centering{}%
\begin{tabular}{rcccc}
 & Processes & Threads/process & SIMD enabled & GPU enabled\tabularnewline
 & (MPI) & (OpenMP) & (SSE) & (CUDA)\tabularnewline
 & \texttt{-{}-enable-mpi -{}-mpi-np }\texttt{\emph{n}} & \texttt{-{}-nthreads }\texttt{\emph{n}} & \texttt{-{}-enable-sse} & \texttt{-{}-enable-cuda}\tabularnewline
\hline 
\#1 & 1 & 1 &  & \tabularnewline
\#2 & 1 & 4 &  & \tabularnewline
\#3 & 1 & 1 & $\checkmark$ & \tabularnewline
\#4 & 1 & 4 & $\checkmark$ & \tabularnewline
\#5 & 1 & 1 &  & $\checkmark$\tabularnewline
\#6 & 1 & 4 &  & $\checkmark$\tabularnewline
\#7 & 3 & 4 & $\checkmark$ & \tabularnewline
\#8 & 3 & 4 &  & $\checkmark$\tabularnewline
\hline 
\end{tabular}\caption{System configurations for timing results. Headings additionally give
the technology (e.g., MPI) and \pkg{LibBi} command-line option (e.g.,
\texttt{-{}-enable-mpi}) used to enable it. All configurations additionally
use the \texttt{-{}-disable-assert} command-line option to disable
runtime assertion checks.\label{tab:system-config}}
\end{table}

\begin{figure}[tp]
\includegraphics[width=1\textwidth]{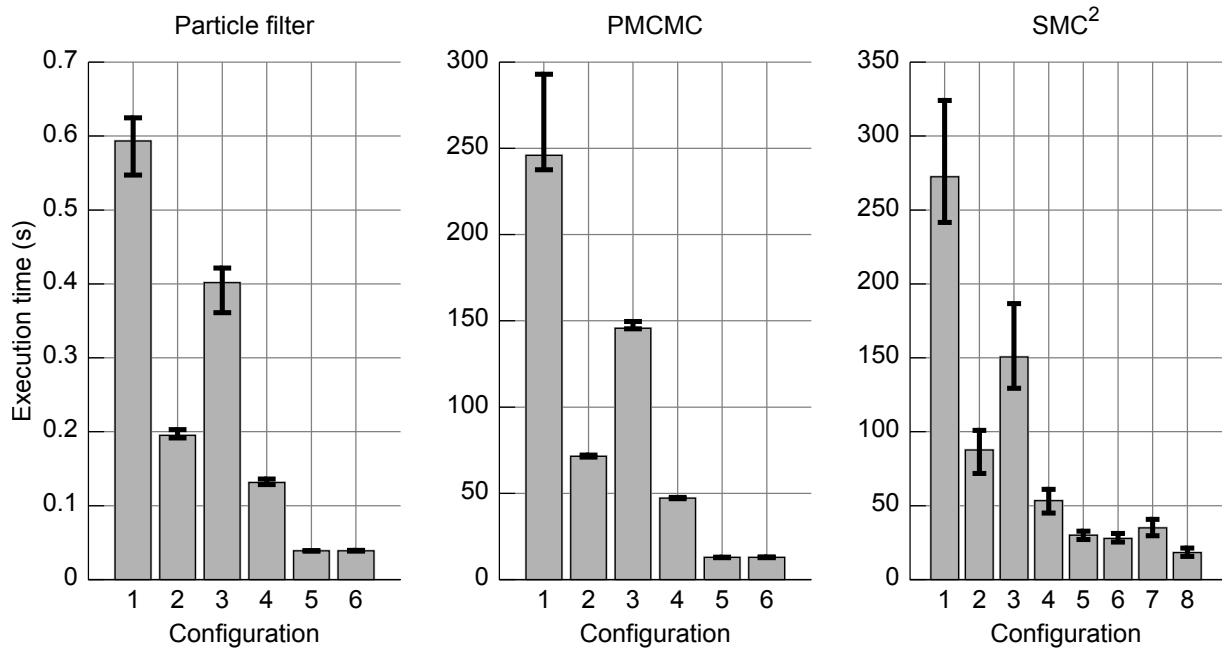}\caption{Execution times for each method under each configuration. Results
are based on 100 runs with different random number seeds. Columns
give the median, while error bars straddle the 50\% credibility interval.\label{fig:time}}
\end{figure}

Figure \ref{fig:time} gives the execution times for each method in
each configuration, summarised across 100 repeated runs with different
random number seeds. For context, one would ideally like to see:
\begin{enumerate}
\item that configurations \#2 and \#4 execute four times faster than configurations
\#1 and \#3, respectively, due to OpenMP multithreading,
\item that configurations \#3 and \#4 execute four times faster than configurations
\#1 and \#2, respectively, due to four-way SSE vector parallelism
in single precision floating point,
\item that configurations \#5 and \#6 execute faster than configuration
\#4, as the algorithms are well-suited to GPU, and
\item that configurations \#7 and \#8 execute three times faster than configurations
\#4 and \#6, respectively, due to three MPI processes on uncontested
hardware.
\end{enumerate}
Of course, these ideal targets are rarely met in practice due to synchronisation
overhead and necessarily serial sections of code. Results in Figure
\ref{fig:time} do suggest good gains with OpenMP, SSE and GPU use,
however, and modest gains in MPI use. Improving gains with MPI is
a topic of current research.

Performance of these methods is very much model-dependent, and the
results of Figure \ref{fig:time} should be considered demonstrative
only. For example, \pkg{LibBi} parallelises predominantly (but not
only, e.g., \citet{Murray2012}) across particles. For GPU performance
to exceed CPU-only performance, typically at least a thousand particles
will be required. The number of $\mathbf{x}$-particles per \textbf{$\boldsymbol{\theta}$}
sample in these configurations (Table \ref{tab:method-config}) is
sufficiently large that good returns are expected. Not all models
justify the use of this many $\mathbf{x}$-particles, however, and
for fewer, CPU performance can, and in many cases does, exceed GPU
performance.

\section{Summary}

\label{sec:summary}

\pkg{LibBi} combines the utility the SSMs with the generality and
computational scalability of SMC methods, in a manner sensitive to
modern hardware realities. Its two aims are accessibility and speed.
Accessibility is achieved by decoupling models, from methods, from
the hardware on which they run. Models are specified in a modelling
language with a rich variety of features, the focus on SMC methods
ensures their broad inferential support, and code generation alleviates
the user from the combinatorial task of writing model-, method- and
hardware-specific code. Speed is achieved through code generation
and template metaprogramming techniques, along with support for high-performance
technologies such as SSE, OpenMP, MPI and CUDA to make full use of
available hardware resources, including GPUs. Good performance gains
are obtained with these technologies as demonstrated in Figure \ref{fig:time}.
Research continues into both statistical and computational advances
to scale to higher-dimensional models on large compute clusters.

\section{Supplementary material}

\label{sec:supplementary}

The two examples given in this work are available for download at
\url{http://www.libbi.org/examples.html}, as the \pkg{Windkessel}
and \pkg{Lorenz96} packages. Each package includes the model, data,
and input files required to reproduce the results in this work.

\bibliographystyle{abbrvnat}

\input{libbi_arxiv.bbl}
\end{document}

%% file: Windkessel.bi.tex
\pagecolor{bgcolor}
\noindent
\ttfamily
\hlstd{}\hllin{\ \ \ \ 1\ }\hlcom{/{*}{*}}\\
\hllin{\ \ \ \ 2\ }\hlcom{\ {*}\ Three{-}element\ Windkessel\ model.}\\
\hllin{\ \ \ \ 3\ }\hlcom{\ {*}/}\hlstd{}\\
\hllin{\ \ \ \ 4\ }\hlkwb{model\ }\hlstd{Windkessel\ }\hlopt{\{}\\
\hllin{\ \ \ \ 5\ }\hlstd{}\hlstd{\ \ }\hlstd{}\hlkwb{const\ }\hlstd{h\ }\hlopt{=\ }\hlstd{}\hlnum{0.01}\hlstd{\ \ }\hlnum{}\hlstd{}\hlslc{//\ time\ step\ (s)}\\
\hllin{\ \ \ \ 6\ }\hlstd{\\
\hllin{\ \ \ \ 7\ }}\hlstd{\ \ }\hlstd{}\hlkwb{param\ }\hlstd{R}\hlstd{\ \ \ \ \ \ \ \ \ }\hlstd{}\hlslc{//\ peripheral\ resistance,\ mm\ Hg\ (ml\ s{*}{*}{-}1){*}{*}{-}1}\\
\hllin{\ \ \ \ 8\ }\hlstd{}\hlstd{\ \ }\hlstd{}\hlkwb{param\ }\hlstd{C}\hlstd{\ \ \ \ \ \ \ \ \ }\hlstd{}\hlslc{//\ arterial\ compliance,\ ml\ (mm\ Hg){*}{*}{-}1}\\
\hllin{\ \ \ \ 9\ }\hlstd{}\hlstd{\ \ }\hlstd{}\hlkwb{param\ }\hlstd{Z}\hlstd{\ \ \ \ \ \ \ \ \ }\hlstd{}\hlslc{//\ characteristic\ impedence,\ mm\ Hg\ s\ ml{*}{*}{-}1}\\
\hllin{\ \ \ 10\ }\hlstd{}\hlstd{\ \ }\hlstd{}\hlkwb{param\ }\hlstd{sigma2}\hlstd{\ \ \ \ }\hlstd{}\hlslc{//\ process\ noise\ variance,\ (mm\ Hg){*}{*}2}\\
\hllin{\ \ \ 11\ }\hlstd{}\hlstd{\ \ }\hlstd{}\hlkwb{input\ }\hlstd{F}\hlstd{\ \ \ \ \ \ \ \ \ }\hlstd{}\hlslc{//\ aortic\ flow,\ ml\ s{*}{*}{-}1}\\
\hllin{\ \ \ 12\ }\hlstd{}\hlstd{\ \ }\hlstd{}\hlkwb{noise\ }\hlstd{xi}\hlstd{\ \ \ \ \ \ \ \ }\hlstd{}\hlslc{//\ noise,\ ml\ s{*}{*}{-}1}\\
\hllin{\ \ \ 13\ }\hlstd{}\hlstd{\ \ }\hlstd{}\hlkwb{state\ }\hlstd{Pp}\hlstd{\ \ \ \ \ \ \ \ }\hlstd{}\hlslc{//\ peripheral\ pressure,\ mm\ Hg}\\
\hllin{\ \ \ 14\ }\hlstd{}\hlstd{\ \ }\hlstd{}\hlkwb{obs\ }\hlstd{Pa}\hlstd{\ \ \ \ \ \ \ \ \ \ }\hlstd{}\hlslc{//\ observed\ aortic\ pressure,\ mm\ Hg}\\
\hllin{\ \ \ 15\ }\hlstd{\\
\hllin{\ \ \ 16\ }}\hlstd{\ \ }\hlstd{}\hlkwb{sub\ }\hlstd{}\hlkwd{parameter\ }\hlstd{}\hlopt{\{}\\
\hllin{\ \ \ 17\ }\hlstd{}\hlstd{\ \ \ \ }\hlstd{R\ }\hlopt{\textasciitilde{} }\hlstd{}\hlkwd{gamma}\hlstd{}\hlopt{(}\hlstd{}\hlnum{2.0}\hlstd{}\hlopt{,\ }\hlstd{}\hlnum{0.9}\hlstd{}\hlopt{)}\\
\hllin{\ \ \ 18\ }\hlstd{}\hlstd{\ \ \ \ }\hlstd{C\ }\hlopt{\textasciitilde{}\ }\hlstd{}\hlkwd{gamma}\hlstd{}\hlopt{(}\hlstd{}\hlnum{2.0}\hlstd{}\hlopt{,\ }\hlstd{}\hlnum{1.5}\hlstd{}\hlopt{)}\\
\hllin{\ \ \ 19\ }\hlstd{}\hlstd{\ \ \ \ }\hlstd{Z\ }\hlopt{\textasciitilde{}\ }\hlstd{}\hlkwd{gamma}\hlstd{}\hlopt{(}\hlstd{}\hlnum{2.0}\hlstd{}\hlopt{,\ }\hlstd{}\hlnum{0.03}\hlstd{}\hlopt{)}\\
\hllin{\ \ \ 20\ }\hlstd{}\hlstd{\ \ \ \ }\hlstd{sigma2\ }\hlopt{\textasciitilde{}\ }\hlstd{}\hlkwd{inverse\textunderscore gamma}\hlstd{}\hlopt{(}\hlstd{}\hlnum{2.0}\hlstd{}\hlopt{,\ }\hlstd{}\hlnum{25.0}\hlstd{}\hlopt{)}\\
\hllin{\ \ \ 21\ }\hlstd{}\hlstd{\ \ }\hlstd{}\hlopt{\}}\\
\hllin{\ \ \ 22\ }\hlstd{\\
\hllin{\ \ \ 23\ }}\hlstd{\ \ }\hlstd{}\hlkwb{sub\ }\hlstd{}\hlkwd{initial\ }\hlstd{}\hlopt{\{}\\
\hllin{\ \ \ 24\ }\hlstd{}\hlstd{\ \ \ \ }\hlstd{Pp\ }\hlopt{\textasciitilde{}\ }\hlstd{}\hlkwd{gaussian}\hlstd{}\hlopt{(}\hlstd{}\hlnum{90.0}\hlstd{}\hlopt{,\ }\hlstd{}\hlnum{15.0}\hlstd{}\hlopt{)}\\
\hllin{\ \ \ 25\ }\hlstd{}\hlstd{\ \ }\hlstd{}\hlopt{\}}\\
\hllin{\ \ \ 26\ }\hlstd{\\
\hllin{\ \ \ 27\ }}\hlstd{\ \ }\hlstd{}\hlkwb{sub\ }\hlstd{}\hlkwd{transition}\hlstd{}\hlopt{(}\hlstd{delta\ }\hlopt{=\ }\hlstd{h}\hlopt{)\ \{}\\
\hllin{\ \ \ 28\ }\hlstd{}\hlstd{\ \ \ \ }\hlstd{xi\ }\hlopt{\textasciitilde{}\ }\hlstd{}\hlkwd{gaussian}\hlstd{}\hlopt{(}\hlstd{}\hlnum{0.0}\hlstd{}\hlopt{,\ }\hlstd{h}\hlopt{{*}}\hlstd{sqrt}\hlopt{(}\hlstd{sigma2}\hlopt{))}\\
\hllin{\ \ \ 29\ }\hlstd{}\hlstd{\ \ \ \ }\hlstd{Pp\ }\hlopt{<-\ }\hlstd{}\hlkwd{exp}\hlstd{}\hlopt{({-}}\hlstd{h}\hlopt{/(}\hlstd{R}\hlopt{{*}}\hlstd{C}\hlopt{)){*}}\hlstd{Pp\ }\hlopt{+\ }\hlstd{R}\hlopt{{*}(}\hlstd{}\hlnum{1.0\ }\hlstd{}\hlopt{{-}\ }\hlstd{exp}\hlopt{({-}}\hlstd{h}\hlopt{/(}\hlstd{R}\hlopt{{*}}\hlstd{C}\hlopt{))){*}(}\hlstd{F\ }\hlopt{+\ }\hlstd{xi}\hlopt{)}\\
\hllin{\ \ \ 30\ }\hlstd{}\hlstd{\ \ }\hlstd{}\hlopt{\}}\\
\hllin{\ \ \ 31\ }\hlstd{\\
\hllin{\ \ \ 32\ }}\hlstd{\ \ }\hlstd{}\hlkwb{sub\ }\hlstd{}\hlkwd{observation\ }\hlstd{}\hlopt{\{}\\
\hllin{\ \ \ 33\ }\hlstd{}\hlstd{\ \ \ \ }\hlstd{Pa\ }\hlopt{\textasciitilde{}\ }\hlstd{}\hlkwd{gaussian}\hlstd{}\hlopt{(}\hlstd{Pp\ }\hlopt{+\ }\hlstd{Z}\hlopt{{*}}\hlstd{F}\hlopt{,\ }\hlstd{}\hlnum{2.0}\hlstd{}\hlopt{)}\\
\hllin{\ \ \ 34\ }\hlstd{}\hlstd{\ \ }\hlstd{}\hlopt{\}}\\
\hllin{\ \ \ 35\ }\hlstd{\\
\hllin{\ \ \ 36\ }}\hlstd{\ \ }\hlstd{}\hlkwb{sub\ }\hlstd{}\hlkwd{proposal\textunderscore parameter\ }\hlstd{}\hlopt{\{}\\
\hllin{\ \ \ 37\ }\hlstd{}\hlstd{\ \ \ \ }\hlstd{R\ }\hlopt{\textasciitilde{}\ }\hlstd{}\hlkwd{truncated\textunderscore gaussian}\hlstd{}\hlopt{(}\hlstd{R}\hlopt{,\ }\hlstd{}\hlnum{0.03}\hlstd{}\hlopt{,\ }\hlstd{lower\ }\hlopt{=\ }\hlstd{}\hlnum{0.0}\hlstd{}\hlopt{)}\\
\hllin{\ \ \ 38\ }\hlstd{}\hlstd{\ \ \ \ }\hlstd{C\ }\hlopt{\textasciitilde{}\ }\hlstd{}\hlkwd{truncated\textunderscore gaussian}\hlstd{}\hlopt{(}\hlstd{C}\hlopt{,\ }\hlstd{}\hlnum{0.1}\hlstd{}\hlopt{,\ }\hlstd{lower\ }\hlopt{=\ }\hlstd{}\hlnum{0.0}\hlstd{}\hlopt{)}\\
\hllin{\ \ \ 39\ }\hlstd{}\hlstd{\ \ \ \ }\hlstd{Z\ }\hlopt{\textasciitilde{}\ }\hlstd{}\hlkwd{truncated\textunderscore gaussian}\hlstd{}\hlopt{(}\hlstd{Z}\hlopt{,\ }\hlstd{}\hlnum{0.002}\hlstd{}\hlopt{,\ }\hlstd{lower\ }\hlopt{=\ }\hlstd{}\hlnum{0.0}\hlstd{}\hlopt{)}\\
\hllin{\ \ \ 40\ }\hlstd{}\hlstd{\ \ \ \ }\hlstd{sigma2\ }\hlopt{\textasciitilde{}\ }\hlstd{}\hlkwd{inverse\textunderscore gamma}\hlstd{}\hlopt{(}\hlstd{}\hlnum{2.0}\hlstd{}\hlopt{,\ }\hlstd{}\hlnum{3.0}\hlstd{}\hlopt{{*}}\hlstd{sigma2}\hlopt{)}\\
\hllin{\ \ \ 41\ }\hlstd{}\hlstd{\ \ }\hlstd{}\hlopt{\}}\\
\hllin{\ \ \ 42\ }\hlstd{}\hlopt{\}}\hlstd{}\\
\mbox{}
\normalfont
\normalsize

%% file: Lorenz96.bi.tex
\noindent
\ttfamily
\hlstd{}\hllin{\ \ \ \ 1\ }\hlcom{/{*}{*}}\\
\hllin{\ \ \ \ 2\ }\hlcom{\ {*}\ Lorenz\ '96\ model.}\\
\hllin{\ \ \ \ 3\ }\hlcom{\ {*}/}\hlstd{}\\
\hllin{\ \ \ \ 4\ }\hlkwb{model\ }\hlstd{Lorenz96\ }\hlopt{\{}\\
\hllin{\ \ \ \ 5\ }\hlstd{}\hlstd{\ \ }\hlstd{}\hlkwb{dim\ }\hlstd{n}\hlopt{(}\hlstd{size\ }\hlopt{=\ }\hlstd{}\hlnum{8}\hlstd{}\hlopt{,\ }\hlstd{boundary\ }\hlopt{=\ }\hlstd{}\hlstr{'cyclic'}\hlstd{}\hlopt{)}\\
\hllin{\ \ \ \ 6\ }\hlstd{\\
\hllin{\ \ \ \ 7\ }}\hlstd{\ \ }\hlstd{}\hlkwb{const\ }\hlstd{h\ }\hlopt{=\ }\hlstd{}\hlnum{0.05}\hlstd{\ \ \ }\hlnum{}\hlstd{}\hlslc{//\ step\ size}\\
\hllin{\ \ \ \ 8\ }\hlstd{\\
\hllin{\ \ \ \ 9\ }}\hlstd{\ \ }\hlstd{}\hlkwb{param\ }\hlstd{F}\hlstd{\ \ \ \ \ \ \ \ \ \ }\hlstd{}\hlslc{//\ forcing}\\
\hllin{\ \ \ 10\ }\hlstd{}\hlstd{\ \ }\hlstd{}\hlkwb{param\ }\hlstd{sigma2}\hlstd{\ \ \ \ \ }\hlstd{}\hlslc{//\ diffusion\ variance}\\
\hllin{\ \ \ 11\ }\hlstd{}\hlstd{\ \ }\hlstd{}\hlkwb{state\ }\hlstd{x}\hlopt{{[}}\hlstd{n}\hlopt{{]}}\hlstd{\ \ \ \ \ \ \ }\hlopt{}\hlstd{}\hlslc{//\ state\ variables}\\
\hllin{\ \ \ 12\ }\hlstd{}\hlstd{\ \ }\hlstd{}\hlkwb{noise\ }\hlstd{deltaW}\hlopt{{[}}\hlstd{n}\hlopt{{]}}\hlstd{\ \ }\hlopt{}\hlstd{}\hlslc{//\ Wiener\ process\ increments}\\
\hllin{\ \ \ 13\ }\hlstd{}\hlstd{\ \ }\hlstd{}\hlkwb{obs\ }\hlstd{y}\hlopt{{[}}\hlstd{n}\hlopt{{]}}\hlstd{\ \ \ \ \ \ \ \ \ }\hlopt{}\hlstd{}\hlslc{//\ observations}\\
\hllin{\ \ \ 14\ }\hlstd{\\
\hllin{\ \ \ 15\ }}\hlstd{\ \ }\hlstd{}\hlkwb{sub\ }\hlstd{}\hlkwd{parameter\ }\hlstd{}\hlopt{\{}\\
\hllin{\ \ \ 16\ }\hlstd{}\hlstd{\ \ \ \ }\hlstd{F\ }\hlopt{\textasciitilde{}\ }\hlstd{}\hlkwd{uniform}\hlstd{}\hlopt{(}\hlstd{}\hlnum{8.0}\hlstd{}\hlopt{,\ }\hlstd{}\hlnum{12.0}\hlstd{}\hlopt{)}\\
\hllin{\ \ \ 17\ }\hlstd{}\hlstd{\ \ \ \ }\hlstd{sigma2\ }\hlopt{\textasciitilde{}\ }\hlstd{}\hlkwd{inverse\textunderscore gamma}\hlstd{}\hlopt{(}\hlstd{}\hlnum{2.0}\hlstd{}\hlopt{,\ }\hlstd{}\hlnum{0.25}\hlstd{}\hlopt{)}\\
\hllin{\ \ \ 18\ }\hlstd{}\hlstd{\ \ }\hlstd{}\hlopt{\}}\\
\hllin{\ \ \ 19\ }\hlstd{\\
\hllin{\ \ \ 20\ }}\hlstd{\ \ }\hlstd{}\hlkwb{sub\ }\hlstd{}\hlkwd{initial\ }\hlstd{}\hlopt{\{}\\
\hllin{\ \ \ 21\ }\hlstd{}\hlstd{\ \ \ \ }\hlstd{x}\hlopt{{[}}\hlstd{n}\hlopt{{]}\ \textasciitilde{}\ }\hlstd{}\hlkwd{uniform}\hlstd{}\hlopt{({-}}\hlstd{}\hlnum{1.0}\hlstd{}\hlopt{,\ }\hlstd{}\hlnum{3.0}\hlstd{}\hlopt{)}\\
\hllin{\ \ \ 22\ }\hlstd{}\hlstd{\ \ }\hlstd{}\hlopt{\}}\\
\hllin{\ \ \ 23\ }\hlstd{\\
\hllin{\ \ \ 24\ }}\hlstd{\ \ }\hlstd{}\hlkwb{sub\ }\hlstd{}\hlkwd{transition}\hlstd{}\hlopt{(}\hlstd{delta\ }\hlopt{=\ }\hlstd{h}\hlopt{)\ \{}\\
\hllin{\ \ \ 25\ }\hlstd{}\hlstd{\ \ \ \ }\hlstd{deltaW}\hlopt{{[}}\hlstd{n}\hlopt{{]}\ \textasciitilde{}\ }\hlstd{}\hlkwd{wiener}\hlstd{}\hlopt{()}\\
\hllin{\ \ \ 26\ }\hlstd{}\hlstd{\ \ \ \ }\hlstd{ode}\hlopt{(}\hlstd{h\ }\hlopt{=\ }\hlstd{h}\hlopt{,\ }\hlstd{alg\ }\hlopt{=\ }\hlstd{}\hlstr{'RK4'}\hlstd{}\hlopt{)\ \{}\\
\hllin{\ \ \ 27\ }\hlstd{}\hlstd{\ \ \ \ \ \ }\hlstd{dx}\hlopt{{[}}\hlstd{n}\hlopt{{]}/}\hlstd{dt\ }\hlopt{=\ }\hlstd{x}\hlopt{{[}}\hlstd{n}\hlopt{{-}}\hlstd{}\hlnum{1}\hlstd{}\hlopt{{]}{*}(}\hlstd{x}\hlopt{{[}}\hlstd{n}\hlopt{+}\hlstd{}\hlnum{1}\hlstd{}\hlopt{{]}\ {-}\ }\hlstd{x}\hlopt{{[}}\hlstd{n}\hlopt{{-}}\hlstd{}\hlnum{2}\hlstd{}\hlopt{{]})\ {-}\ }\hlstd{x}\hlopt{{[}}\hlstd{n}\hlopt{{]}\ +\ }\hlstd{F\ }\hlopt{+\ }\hlstd{sqrt}\hlopt{(}\hlstd{sigma2}\hlopt{){*}}\hlstd{deltaW}\hlopt{{[}}\hlstd{n}\hlopt{{]}/}\hlstd{h\\
\hllin{\ \ \ 28\ }}\hlstd{\ \ \ \ }\hlstd{}\hlopt{\}}\\
\hllin{\ \ \ 29\ }\hlstd{}\hlstd{\ \ }\hlstd{}\hlopt{\}}\\
\hllin{\ \ \ 30\ }\hlstd{\\
\hllin{\ \ \ 31\ }}\hlstd{\ \ }\hlstd{}\hlkwb{sub\ }\hlstd{}\hlkwd{observation\ }\hlstd{}\hlopt{\{}\\
\hllin{\ \ \ 32\ }\hlstd{}\hlstd{\ \ \ \ }\hlstd{y}\hlopt{{[}}\hlstd{n}\hlopt{{]}\ \textasciitilde{}\ }\hlstd{}\hlkwd{normal}\hlstd{}\hlopt{(}\hlstd{x}\hlopt{{[}}\hlstd{n}\hlopt{{]},\ }\hlstd{}\hlnum{0.5}\hlstd{}\hlopt{)}\\
\hllin{\ \ \ 33\ }\hlstd{}\hlstd{\ \ }\hlstd{}\hlopt{\}}\\
\hllin{\ \ \ 34\ }\hlstd{\\
\hllin{\ \ \ 35\ }}\hlstd{\ \ }\hlstd{}\hlkwb{sub\ }\hlstd{}\hlkwd{proposal\textunderscore parameter\ }\hlstd{}\hlopt{\{}\\
\hllin{\ \ \ 36\ }\hlstd{}\hlstd{\ \ \ \ }\hlstd{F\ }\hlopt{\textasciitilde{}\ }\hlstd{}\hlkwd{truncated\textunderscore gaussian}\hlstd{}\hlopt{(}\hlstd{F}\hlopt{,\ }\hlstd{}\hlnum{0.1}\hlstd{}\hlopt{,\ }\hlstd{}\hlnum{8.0}\hlstd{}\hlopt{,\ }\hlstd{}\hlnum{12.0}\hlstd{}\hlopt{);}\\
\hllin{\ \ \ 37\ }\hlstd{}\hlstd{\ \ \ \ }\hlstd{sigma2\ }\hlopt{\textasciitilde{}\ }\hlstd{}\hlkwd{inverse\textunderscore gamma}\hlstd{}\hlopt{(}\hlstd{}\hlnum{2.0}\hlstd{}\hlopt{,\ }\hlstd{}\hlnum{3.0}\hlstd{}\hlopt{{*}}\hlstd{sigma2}\hlopt{)}\\
\hllin{\ \ \ 38\ }\hlstd{}\hlstd{\ \ }\hlstd{}\hlopt{\}}\\
\hllin{\ \ \ 39\ }\hlstd{\\
\hllin{\ \ \ 40\ }}\hlstd{\ \ }\hlstd{}\hlkwb{sub\ }\hlstd{}\hlkwd{proposal\textunderscore initial\ }\hlstd{}\hlopt{\{}\\
\hllin{\ \ \ 41\ }\hlstd{}\hlstd{\ \ \ \ }\hlstd{x}\hlopt{{[}}\hlstd{n}\hlopt{{]}\ \textasciitilde{}\ }\hlstd{}\hlkwd{truncated\textunderscore gaussian}\hlstd{}\hlopt{(}\hlstd{x}\hlopt{{[}}\hlstd{n}\hlopt{{]},\ }\hlstd{}\hlnum{0.1}\hlstd{}\hlopt{,\ {-}}\hlstd{}\hlnum{1.0}\hlstd{}\hlopt{,\ }\hlstd{}\hlnum{3.0}\hlstd{}\hlopt{)}\\
\hllin{\ \ \ 42\ }\hlstd{}\hlstd{\ \ }\hlstd{}\hlopt{\}}\\
\hllin{\ \ \ 43\ }\hlstd{}\hlopt{\}}\hlstd{}\\
\mbox{}
\normalfont
\normalsize

%% file: kalman-filter.tex
\begin{codebox}
\Procname{$\proc{Kalman-Filter}(\boldsymbol{\theta}, t_{r:s}) \rightarrow (l,
  \mathbf{x}'(t_{0:s}))$}
\li \If $r = 0$
\li   initialise $\mathbf{X}(t_0) \sim \textsc{N}(\boldsymbol{\mu}(t_0)$, $\mathbf{U}(t_0))$
    \End
\li \For $i = r + 1,\ldots,s$
\zi   \Comment state prediction
\li   $\hat{\boldsymbol{\mu}}(t_i) \leftarrow \mathbf{F}_{\boldsymbol{\theta}}(t_i)\boldsymbol{\mu}(t_{i-1})$
\li   $\mathbf{C}(t_i) \leftarrow \mathbf{U}(t_{i-1})^{\top}\mathbf{U}(t_{i-1})\mathbf{F}_{\boldsymbol{\theta}}(t_i)$
\li   $\hat{\boldsymbol{\Sigma}}(t_i) \leftarrow
\mathbf{F}_{\boldsymbol{\theta}}(t_i)^{\top}\mathbf{U}(t_{i-1})^{\top}\mathbf{U}(t_{i-1})\mathbf{F}_{\boldsymbol{\theta}}(t_i) +
\mathbf{Q}_{\boldsymbol{\theta}}(t_i)^{\top}\mathbf{Q}_{\boldsymbol{\theta}}(t_i)$
\li   $\hat{\mathbf{U}}(t_i) \leftarrow \proc{Cholesky-Factorise}(\hat{\boldsymbol{\Sigma}}(t_i))$
\zi
\zi   \Comment observation prediction
\li   $\boldsymbol{\nu}(t_i) \leftarrow \mathbf{G}_{\boldsymbol{\theta}}(t_i)\hat{\boldsymbol{\mu}}(t_i)$
\li   $\mathbf{D}(t_i) \leftarrow
\hat{\mathbf{U}}(t_i)^{\top}\hat{\mathbf{U}}(t_i)\mathbf{G}_{\boldsymbol{\theta}}(t_i)$
\li   $\mathbf{T}(t_i) \leftarrow
\mathbf{G}_{\boldsymbol{\theta}}(t_i)^{\top}\hat{\mathbf{U}}(t_i)\hat{\mathbf{U}}(t_i)^{\top}\mathbf{G}_{\boldsymbol{\theta}}(t_i) +
\mathbf{R}_{\boldsymbol{\theta}}(t_i)^{\top}\mathbf{R}_{\boldsymbol{\theta}}(t_i)$
\li   $\mathbf{V}(t_i) \leftarrow \proc{Cholesky-Factorise}(\mathbf{T}(t_i))$
\zi
\zi   \Comment correction
\li   $\mathbf{K} \leftarrow \mathbf{D}(t_i)\mathbf{V}(t_i)^{-1}$
\li   $\boldsymbol{\mu}(t_i) \leftarrow \hat{\boldsymbol{\mu}}(t_i) +
\mathbf{K}\mathbf{V}(t_i)^{-\top}(\mathbf{y}(t_i) - \boldsymbol{\nu}(t_i))$
\li   $\mathbf{U}(t_i) \leftarrow \proc{Cholesky-Downdate}(\hat{\mathbf{U}}(t_i), \mathbf{K})$
    \End
\zi
\li $l \leftarrow
\prod_{i=1}^s\frac{1}{\sqrt{2\pi}|\mathbf{V}(t_i)|}\exp\left(-\frac{1}{2}\left\Vert
\mathbf{V}^{-1}(t_i)\left(\mathbf{y}(t_i)-\boldsymbol{\nu}(t_i)\right)\right\Vert^{2}\right)$
\Comment marginal likelihood \label{line:kf-likelihood}
\zi
\zi \Comment single state sample
\li $\mathbf{x}'(t_s) \sim \textsc{N}(\boldsymbol{\mu}(t_s), \mathbf{U}(t_s))$
\li \For $i = s-1,\ldots,0$
\li   $\mathbf{K} \leftarrow \mathbf{C}(t_{i+1})\hat{\mathbf{U}}(t_{i+1})^{-1}$
\li   $\boldsymbol{\omega} \leftarrow \boldsymbol{\mu}(t_i) +
\mathbf{K}\hat{\mathbf{U}}(t_{i+1})^{-\top}(\mathbf{x}'(t_{i+1}) - \hat{\boldsymbol{\mu}}(t_{i+1}))$
\li   $\mathbf{W} \leftarrow
\proc{Cholesky-Factorise}(\mathbf{U}(t_i)^{\top}\mathbf{U}(t_i) -
\mathbf{K}^{\top}\mathbf{K})$
\li   $\mathbf{x}'(t_i) \sim \textsc{N}(\boldsymbol{\omega}, \mathbf{W})$
    \End
\zi
\li \Return $(l, \mathbf{x}'(t_{0:s}))$
\end{codebox}

%% file: particle-filter.tex
\begin{codebox}
\Procname{$\proc{Particle-Filter}(\boldsymbol{\theta}, t_{r:s}) \rightarrow
  (\hat{l}, \hat{\mathbf{x}}'(t_{0:s}))$}
\li \If $r = 0$
\li   \Foreach $j \in \{1,\ldots,P_x\}$ \label{line:pf-init}
\li     $\mathbf{x}^j(t_0) \sim p(\mathbf{X}_0|\boldsymbol{\theta})$ \Comment initialise particle $j$
\li     $w^j(t_0) \leftarrow 1/P_x$ \Comment initialise weight $j$
      \End
    \End
\li \For $i = r + 1,\ldots,s$ \label{line:pf-time}
\li   \Foreach $j \in \{1,\ldots,P_x\}$ \label{line:pf-particle}
\li     $a^j(t_i) \sim \textsc{Multinomial}(\mathbf{w}(t_{i-1}))$
\Comment ancestor for particle $j$
\li     $\mathbf{x}^j(t_i) \sim p(\mathbf{X}(t_i)|\mathbf{x}^{a^j(t_i)}(t_{i-1}),\boldsymbol{\theta})$
\Comment propagate particle $j$
\li     $w^j(t_i) \leftarrow p(\mathbf{y}(t_i)|\mathbf{x}^j(t_i),\boldsymbol{\theta})$ \Comment weight
particle $j$
      \End
    \End
\li $\hat{l} \leftarrow
\prod_{i=1}^{s}\left(\frac{1}{P_{x}}\sum_{j=1}^{P_{x}}w^{j}(t_{i})\right)$
\Comment marginal likelihood \label{line:pf-likelihood}
\zi
\zi \Comment single state sample
\li $j \sim \textsc{Multinomial}(\mathbf{w}(t_s))$
\li $\hat{\mathbf{x}}'(t_s) \leftarrow \mathbf{x}^j(t_s)$
\li \For $i = s,\ldots,1$
\li   $j \leftarrow a^j(t_i)$
\li   $\hat{\mathbf{x}}'(t_{i-1}) \leftarrow \mathbf{x}^j(t_{i-1})$
    \End
\zi
\li \Return $(\hat{l}, \hat{\mathbf{x}}'(t_{0:s}))$
\end{codebox}

%% file: pmmh.tex
\begin{codebox}
\Procname{$\proc{Marginal-Metropolis-Hastings}(\boldsymbol{\theta}, \mathbf{x}(t_{0:T}), l) \rightarrow (\boldsymbol{\theta}, \mathbf{x}(t_{0:T}), l)$}
\li   $\boldsymbol{\theta}' \sim q(\boldsymbol{\theta}'|\boldsymbol{\theta})$ \Comment propose
\li   $(l',\mathbf{x}'(t_{0:T})) \leftarrow
\proc{Filter}(\boldsymbol{\theta}', t_{0:T})$ \Comment likelihood and state
sample from filter
\li   $\alpha \sim \textsc{U}(0,1)$
\li   \If $\alpha \leq
\frac{l' p(\boldsymbol{\theta}') q(\boldsymbol{\theta}|\boldsymbol{\theta}')}
{l p(\boldsymbol{\theta}) q(\boldsymbol{\theta}'|\boldsymbol{\theta})}$
\li     \Return $(\boldsymbol{\theta}', \mathbf{x}'(t_{0:T}), l')$ \Comment accept
\li   \Else
\li     \Return $(\boldsymbol{\theta}, \mathbf{x}(t_{0:T}), l)$ \Comment reject
      \End
\end{codebox}

%% file: smc2.tex
\begin{codebox}
\Procname{$\proc{Sequential-Monte-Carlo}$}
\li \Foreach $j \in \{1,\ldots,P_\theta\}$ \label{line:smc2-init}
\li   $\boldsymbol{\theta}^j(t_0) \sim p(\boldsymbol{\theta})$ \Comment initialise $\theta$-particle $j$
\li   $v^j(t_0) \leftarrow 1/P_\theta$ \Comment initialise $\theta$-weight $j$
\li   $\mathbf{x}^j(t_0) \sim p(\mathbf{X}_0|\boldsymbol{\theta}^j(t_0))$
\Comment initialise state sample $j$
\li   $l^j(t_0) \leftarrow 1$ \Comment initialise likelihood $j$
    \End
\li \For $i = 1,\ldots,T$
\li   \Foreach $j \in \{1,\ldots,P_\theta\}$ \label{line:smc2-particle}
\li     $a^j(t_i) \sim \textsc{Multinomial}(\mathbf{v}(t_{i-1}))$ \Comment
ancestor for $\theta$-particle $j$
\li     $(\boldsymbol{\theta}^j(t_i),\mathbf{x}^j(t_{0:i-1}),l^j(t_{i-1}))
\leftarrow \proc{Marginal-Metropolis-Hastings}($
\zi     \hspace{2cm} $\boldsymbol{\theta}^{a^j(t_i)}(t_{i-1}),
\mathbf{x}^{a^j(t_i)}(t_{0:i-1}), l^{a^j(t_i)}(t_{i-1}))$ \Comment rejuvenate $\theta$-particle $j$
\li     $(v^j(t_i), \mathbf{x}^j(t_{0:i})) \leftarrow \proc{Filter}(\boldsymbol{\theta}^j(t_i),
t_{i-1:i})$ \Comment propagate and weight $\theta$-particle $j$
      \End
    \End
\end{codebox}

%% file: hardware.tex
\begin{tabular}[width=\textwidth]{lrr}
 & Intel Xeon E5-2650 CPU & Nvidia Tesla S2050 GPU \\
\hline
Task parallism & 8-way & 14-way \\
Data parallism (per task) & 8-way$^*$ & 32-way$^\dag$ \\
Clock rate & 2.00 GHz & 1.15 GHz \\
Cache & 20480 KB & 768 KB \\
\hline
\end{tabular}

\footnotesize{\begin{itemize}
\item[$*$] Using single-precision AVX instructions,
  4-way for double precision. Earlier generation CPUs can use SSE
  instructions, 4-way for single and 2-way for double precision, and indeed
  \pkg{LibBi} is limited to this at time of writing.
\item[$\dag$] Threads per warp.
\end{itemize}}